\documentclass[sigconf,]{acmart}

\renewcommand\footnotetextcopyrightpermission[1]{} %
\setcopyright{none}

\usepackage{lineno}
\usepackage{configures/fancyhdr}
\usepackage{booktabs} 

\usepackage{amssymb}  

\usepackage{array}    
\usepackage{amsmath}

\usepackage{textcomp}
\usepackage{listings}
\usepackage{booktabs}
\usepackage{xspace}
\usepackage{comment}
\usepackage{pifont}

\usepackage{multirow}
\usepackage{algorithm}
\usepackage{algpseudocode}
\usepackage{url}

\usepackage{listings}
\usepackage[font=footnotesize]{caption}
\usepackage{subcaption}
\usepackage[USenglish]{babel}
\usepackage{amsmath}
\usepackage{url}
\usepackage{tabularx}
\usepackage{amsfonts}
\usepackage{tikz}
\usepackage{balance}
\usepackage{url}
\usepackage{soul}
\usepackage{enumitem}

\usepackage{relsize}

\definecolor{codegreen}{rgb}{0,0.6,0}
\definecolor{codegray}{rgb}{0.5,0.5,0.5}
\definecolor{codepurple}{rgb}{0.58,0,0.82}
\definecolor{backcolour}{rgb}{0.98,0.98,0.95}

\lstdefinestyle{mystyle}{
    backgroundcolor=\color{backcolour},
    commentstyle=\color{codegreen},
    keywordstyle=\color{magenta},
    numberstyle=\tiny\color{codegray},
    stringstyle=\color{codepurple},
    basicstyle=\ttfamily\footnotesize,
    breakatwhitespace=false,
    breaklines=true,
    captionpos=b,
    keepspaces=true,
    numbers=left,
    numbersep=5pt,
    showspaces=false,
    showstringspaces=false,
    showtabs=false,
    tabsize=2
}
\newcommand{\SystemName}{\textsc{SpecBox}\xspace}

\definecolor{Gray}{gray}{0.95}

\newcommand\cparagraph[1]{\vspace{1.5mm}\noindent\textbf{#1.}}

\newcolumntype{R}{>{\raggedleft\arraybackslash\hsize=.5\hsize\linewidth=\hsize}X}
\newcolumntype{T}{>{\hsize=2.5\hsize\linewidth=2.5\hsize}X}

\newcommand{\mypara}[1]{{\vspace{0.1mm}\noindent\textbf{#1}}}

\begin{document}
\nolinenumbers
\title{\SystemName: Speculative Sandbox Scheduling for Efficient LLM Agent Serving}

\author{Yihui Zhang\textsuperscript{1}, 
Tianyu Wo\textsuperscript{1}, 
Jinghao Wang\textsuperscript{1},
Xiaoyang Sun\textsuperscript{2},
Menghao Zhang\textsuperscript{1},  
Cangzhou Yuan\textsuperscript{1}, 
Li Li\textsuperscript{1},
Chunming Hu\textsuperscript{1},
Albert Y. Zomaya\textsuperscript{3}, 
Renyu Yang\textsuperscript{1\dag}
\thanks{Corresponding Author: Renyu Yang (renyuyang@buaa.edu.cn)}
}

\affiliation{%
$^1$Beihang University \quad \quad $^2$University of Leeds \quad \quad $^3$The University of Sydney 
\country{}
}

\renewcommand{\authors}{Yihui Zhang, Tianyu Wo, Jinghao Wang, Xiaoyang Sun, Menghao Zhang, Cangzhou Yuan, Li Li, Chunming Hu, Albert Y. Zomaya, Renyu Yang}

\renewcommand{\shortauthors}{Y. Zhang, Tianyu Wo,  et. al}

\begin{abstract}
As LLM agents increasingly rely on the Model Context Protocol (MCP) to invoke isolated external sandboxes, disaggregated sandbox deployment introduces a fundamental tension between resource utilization and interactive tail latency. Persistent long-lived sandbox reservations incur excessive memory overhead at scale, while lazy on-demand instantiation generates severe cold-start penalties that degrade response performance under multi-tenant, multi-turn agent workloads. To resolve this dilemma, we present \SystemName, a runtime built around speculative sandbox preallocation tailored for dynamic LLM agent execution pipelines.

At its core, \SystemName implements keyword matching and streaming semantic embedding to enable intent-driven sandbox prewarming, which identifies pending tool execution demands mid-LLM token generation and fully overlaps sandbox bootstrapping with model inference. To extend prewarming windows across sequential agent steps, the framework leverages context-aware stochastic prefetching atop a sandbox dependency graph to probabilistically forecast future sandbox switches ahead of execution. We complement these speculative mechanisms with two orthogonal optimizations: a semantic result cache that prunes redundant repeated sandbox invocations, and a dedicated out-of-band shared-memory transport plane that bypasses conventional network serialization to deliver zero-copy artifact transfers. Evaluated on high-concurrency multi-turn agent traces, our prototype demonstrates that \SystemName cuts P99 end-to-end latency by up to $2.9\times$ relative to the on-demand sandbox baseline, while slashing peak memory consumption by $45.9\%$ compared to permanently reserved sandbox deployments.
\end{abstract}



\maketitle
\section{Introduction}

Modern LLM agents are evolving from conventional text generators into autonomous systems that iteratively reason, plan, and execute external actions~\cite{yao2022react,li2024survey,wang2024survey}. Unlike traditional LLM serving workloads that primarily optimize token generation~\cite{agrawal2024taming,zhang2025cauchy}, agent execution forms a stateful run-loop where an LLM-based controller repeatedly invokes external tools~\cite{playwright,githubmcp,jupyter,neo4j} such as code execution, web automation, and data processing services.
For security, isolation, and reproducibility, these tools are increasingly deployed as independent sandbox environments and accessed through standardized interfaces such as the Model Context Protocol (MCP)~\cite{mcp2024}. Consequently, the latency of agent execution is no longer determined solely by model inference, but also by the efficiency of coordinating heterogeneous external execution environments.


Modern cloud-native infrastructures increasingly adopt serverless sandbox execution to support large-scale concurrent agent sessions~\cite{aws2026, gcp2026, azure2026}, i.e., the execution environments are initialized on demand rather than being reserved as long-lived instances. This design enables the required elasticity for multi-tenant agent workloads but introduces non-negligible latency \cite{puliafito2022stateless, yu2024rainbowcake,xu2024faasmem, stojkovic2023mxfaas}. Each tool invocation may incur sandbox initialization overhead --- such as image loading, filesystem preparation, namespace configuration, and runtime handshake --- introducing a second-level delay before execution begins~\cite{10.1145/3805475}.
In multi-turn agent workflows, such startup overheads accumulate across successive tool invocations, significantly degrading end-to-end responsiveness and limiting the practicality of interactive agent services.

\begin{figure}[t]
    \centering
    \includegraphics[width=\linewidth]{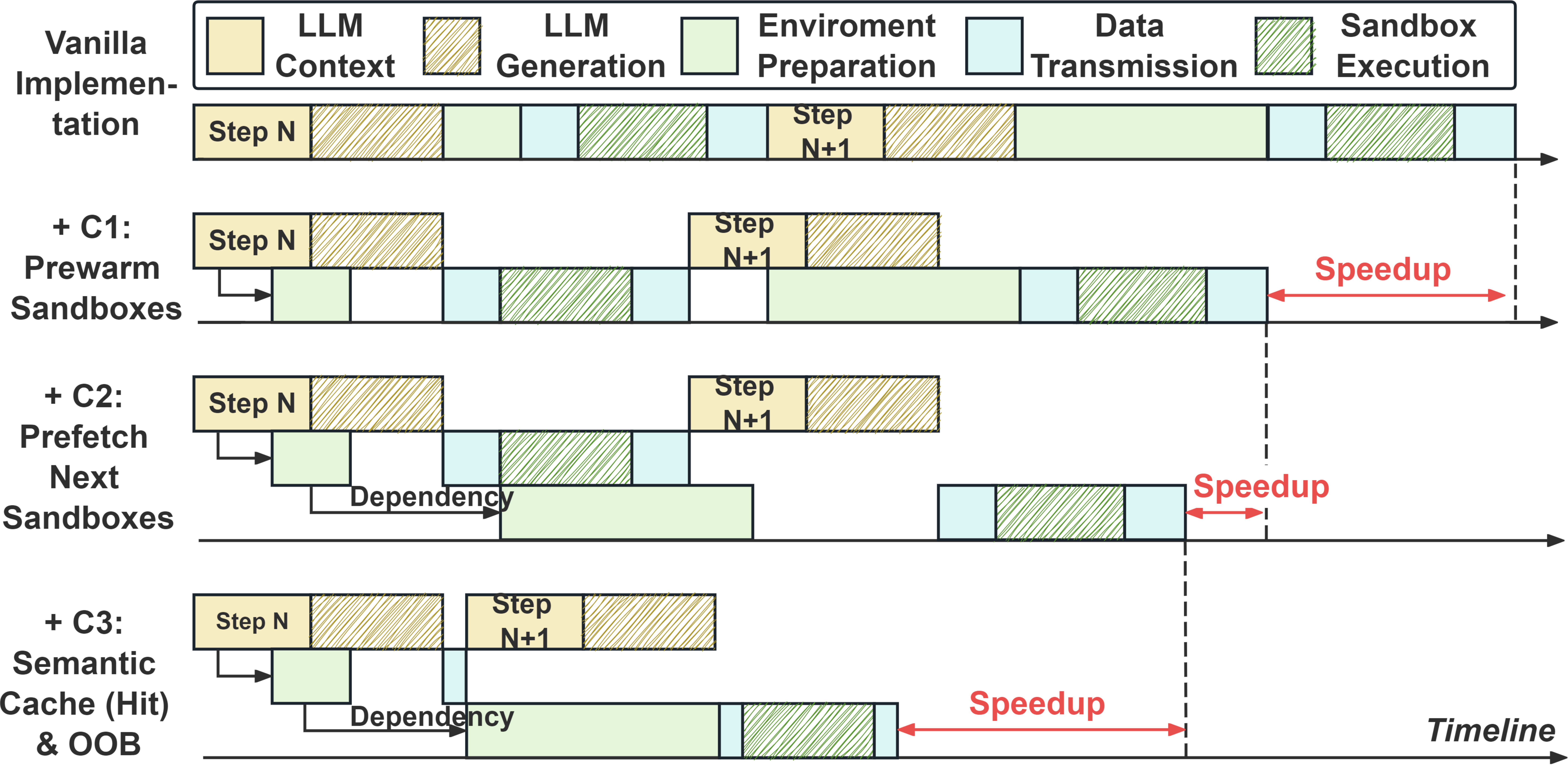}
    \vspace{-1.2em}
    \caption{Proactive and overlapped execution v.s. the native vanilla approaches}
    \vspace{-4mm}
\label{fig:motivation}
\end{figure}

The main cause of this latency is not sandbox initialization, but the sequential execution model used by the vanilla implementation of the de facto agent runtimes (e.g., AgentScope\cite{gao2024agentscope}, AutoGen\cite{wu2023autogen}, and LangGraph~\cite{langgraph}). These systems use a reactive and sequential model without pipelined orchestration: the stage of sandbox preparation begins only after the stage of LLM inference finishes the token generation, and the tool invocation is fully determined. As a result, the preparation of the environment remains entirely exposed on the critical path, leaving idle periods between model reasoning and tool execution. While GPUs are occupied generating tokens, CPU-side resources remain underutilized --- once sandbox initialization begins, the GPU has to wait for external execution to complete. As illustrated in Fig.~\ref{fig:motivation}, the reactive model serializes LLM reasoning and environment preparation, without effectively overlapping computation with environment provisioning.

Existing runtime optimizations such as speculative execution~\cite{stojkovic2023specfaas, li2024distributed}, prewarming~\cite{mahgoub2022orion, sui2024pre}, workflow orchestration~\cite{liu2023doing,li2023dataflower}, and caching~\cite{xu2024faasmem,lu2024serialization} are well studied in serverless and cloud systems. However, they do not directly transfer to LLM agent runtimes because agent execution differs fundamentally from conventional request processing. Prior work usually assumes fixed execution targets or pre-defined workflow graph before runtime. In contrast, autonomous agent workflows are produced online via auto-regressive reasoning: tool calls emerge progressively from streaming token generation and remain uncertain until enough semantic context is available. Multi-turn sessions further evolve based on intermediate observations and outcomes, making future tool choices both workflow-dependent and highly dynamic. These characteristics make simple extensions of existing techniques infeasible.  

In the context of LLM agent runtimes, the key idea is therefore to pre-launch the most likely sandbox before it is requested, overlapping environment preparation with ongoing LLM generation. However, realizing this idea for autonomous agent systems introduces three unique challenges.
First, within a single step (agent iteration), intents must be inferred from early token streams with only partial semantics; using short contexts with low thresholds or broad candidate sets improves overlap but risks false-positive prewarming and wasted memory/CPU. Second, across steps, predicting which tool will be invoked in future steps becomes increasingly unreliable as the prediction window extends; prewarming all plausible successors would revert to reserved-deployment costs. Third, prewarming alone removes only sandbox startup delay: repeated tool executions and large artifact transfers can still remain on the critical path even when a sandbox is ready. The system must therefore reuse semantically equivalent execution results and decouple bulk data transfer from control signaling, while preserving protocol compatibility and user-visible semantics.


This paper presents \SystemName, a predictive serving system that orchestrates tasks of an agent workflow efficiently. \SystemName overlaps LLM execution with environment preparation, breaking the rigid sequential dependencies in vanilla implementation of agent runtimes.
\SystemName diminish end-to-end latency through three synergetic techniques: i) \textit{intent-aware sandbox prewarming} that speculates execution intents on the basis of streaming token outputs, at a proper time, and overlap sandbox preparation with ongoing generation within a step; ii) \textit{stochastic sandbox prefetching}, leveraging historical agent execution traces to anticipate future sandbox needs and prepares the sandboxes for the next step during the current step, thereby lowering cold-start latency; and iii) \textit{reuse-aware data transmission} that exploits semantic similarity to bypass redundant tool execution through semantic caching and decouples large artifact delivery from control signaling through an out-of-band data path, eliminating unnecessary computation and serialization overhead. \SystemName's prototype is implemented and integrated with the open-source AgentScope framework~\cite{gao2024agentscope}. \SystemName is framework-agnostic and highly extensible to any LLM agent serving infrastructures. On diverse multi-turn, high-concurrency agent benchmarks, \SystemName outperforms state-of-the-art serverless runtimes while preserving workflow correctness and protocol compatibility. \SystemName reduces P99 latency by up to $2.9\times$ and peak host memory usage by $45.9\%$.
This paper makes the following key contributions.

\begin{itemize}[leftmargin=1.2em]
    \item characterizing the latency-critical path of multi-turn LLM agent execution and proposing a new sandbox prewarming approach (\textbf{C1}) via inferring streaming token-level intents such that sandbox preparation can be better overlapped with the ongoing LLM inference (\S~\ref{sec:appr:intent_routing}).
    \item devising stochastic sandbox prefetching mechanism (\textbf{C2}) to reduce cross-step cold-start latency by mining historical agent execution traces and pre-warming likely sandbox environments over a sandbox dependency graph (\S~\ref{sec:appr:sandbox_prefetching}).
    \item introducing reuse-aware data transmission (\textbf{C3}), via out-of-band transport and semantic caching, efficiently transferring and reusing intermediate execution artifacts while eliminating redundant sandbox initialization (\S~\ref{sec:appr:reuse_aware_transmission}).
\end{itemize}

\section{Background and Motivation}
\label{sec:bg}

\subsection{Agent Workflow and Environment}
Large Language Model (LLM) applications have shifted from monolithic, single-turn chatbots to distributed autonomous agents. Traditional workflows rely on manually predefined execution graphs, where the operation sequence is fixed before runtime. In contrast, modern \textit{agent workflows} integrate autonomous decision-making: given a high-level objective, an agent can dynamically decompose tasks, select execution sandboxes or external environments, and adapt its trajectory based on intermediate observations. As a result, the workflow is no longer statically specified but instead emerges from continuous interactions between LLM reasoning and external environments~\cite{yao2022react, li2024survey, wang2024survey}.

An agent workflow is realized through a sequence of event-driven execution steps, where each step represents an intermediate task that involves reasoning, planning, or invoking local and remote sandboxes. Tasks that require file system interaction, code execution, web access, or queries against proprietary databases are delegated to isolated external execution layers (\textit{environment} or \textit{tool sandboxes}). Standardized protocols—most notably the Model Context Protocol (MCP)~\cite{mcp2024} proposed by Anthropic—formalize this interface boundary. MCP specifies a common communication layer over transports such as HTTP and Server-Sent Events (SSE), thereby transforming tool invocation into a set of loosely coupled microservices. This architecture is closely aligned with emerging agent-oriented operating systems that conceptualize LLMs as central processing units and external environments as peripheral devices~\cite{meiaios, gao2024agentscope}. From a systems-engineering standpoint, an agent’s execution increasingly manifests as a continuous stream of heterogeneous, RPC-like interactions between the LLM inference core and multiple environment services, rather than as a monolithic, localized compute workload.

\vspace{-0.4em}
\subsection{Serving Runtimes}

Disaggregating Agent Engines from their underlying execution environments enables substantial scalability and architectural flexibility. However, deploying these decoupled components within cloud-native, multi-tenant cluster infrastructures introduces significant challenges related to performance and resource management. Service providers must carefully balance isolation guarantees, resource efficiency, and end-to-end latency, thereby exposing an inherent trade-off between long-running (reserved) and serverless (on-demand) serving paradigms. 

\cparagraph{Reserved Agent Runtime} A straightforward strategy~\cite{tan2025towards,10.1145/3805475} is to maintain pre-initialized, always-on containers for each user tenant and its associated sandboxes. However, in contemporary agent ecosystems comprising thousands of fine-grained tools, sustaining all execution environments in a warmed state becomes prohibitively resource-intensive: idle containers incur substantial host memory and CPU overheads~\cite{xu2024faasmem, stojkovic2023mxfaas}, which in turn lead to pronounced cluster underutilization and resource interference, thereby rendering large-scale multi-tenant deployments economically unsustainable.

\cparagraph{On-demand Agent Runtime} To enhance resource utilization, modern cloud-native platforms commonly employ on-demand runtime provisioning~\cite{du2020catalyzer, shahrad2020serverless}, wherein tool sandboxes are instantiated (``cold-started'') only upon explicit requests from agents. However, this design choice introduces substantial tail latency: the on-demand initialization of an isolated container incurs a series of serialized overheads, including container image download and extraction, network namespace setup, virtual file system mounting, and application-level handshake procedures. Under bursty workloads or in complex multi-turn workflows, these multi-second cold-start delays accumulate along the agent's end-to-end critical path, thereby inflating P99 tail latency and degrading the interactive quality of service (QoS) of real-time intelligent agents.

\subsection{Execution Bottlenecks in Agent Runtimes}

In contrast to conventional LLM serving, an agent session constitutes a continuous, stateful execution loop composed of multiple iterative reasoning and sandboxed execution steps. Prevailing agent frameworks~\cite{gao2024agentscope,wu2023autogen,langgraph} employ a reactive execution paradigm, in which each tool invocation is initiated only after the LLM has completed its reasoning step. This design induces a strictly serialized dependency chain between model inference and environment interaction. The resulting execution workflow is depicted in Fig.~\ref{fig:motivation}.

\begin{figure}[t]
    \centering
    \begin{subfigure}[b]{0.49\columnwidth}
        \centering
        \includegraphics[width=\linewidth]{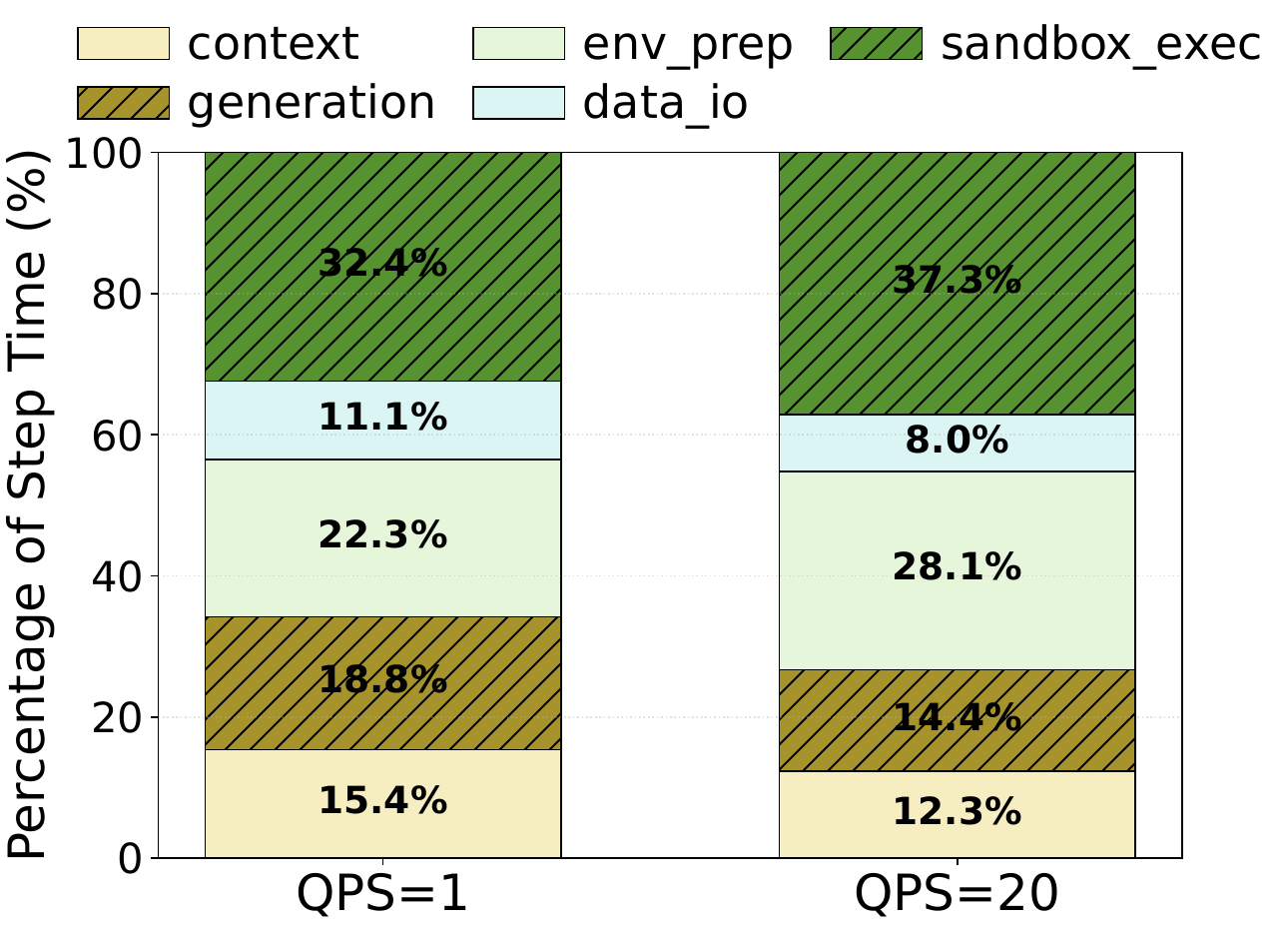}
        \caption{Step time breakdown.}
        \label{fig:time_breakdown}
    \end{subfigure}\hfill
    \begin{subfigure}[b]{0.49\columnwidth}
        \centering
        \includegraphics[width=\linewidth]{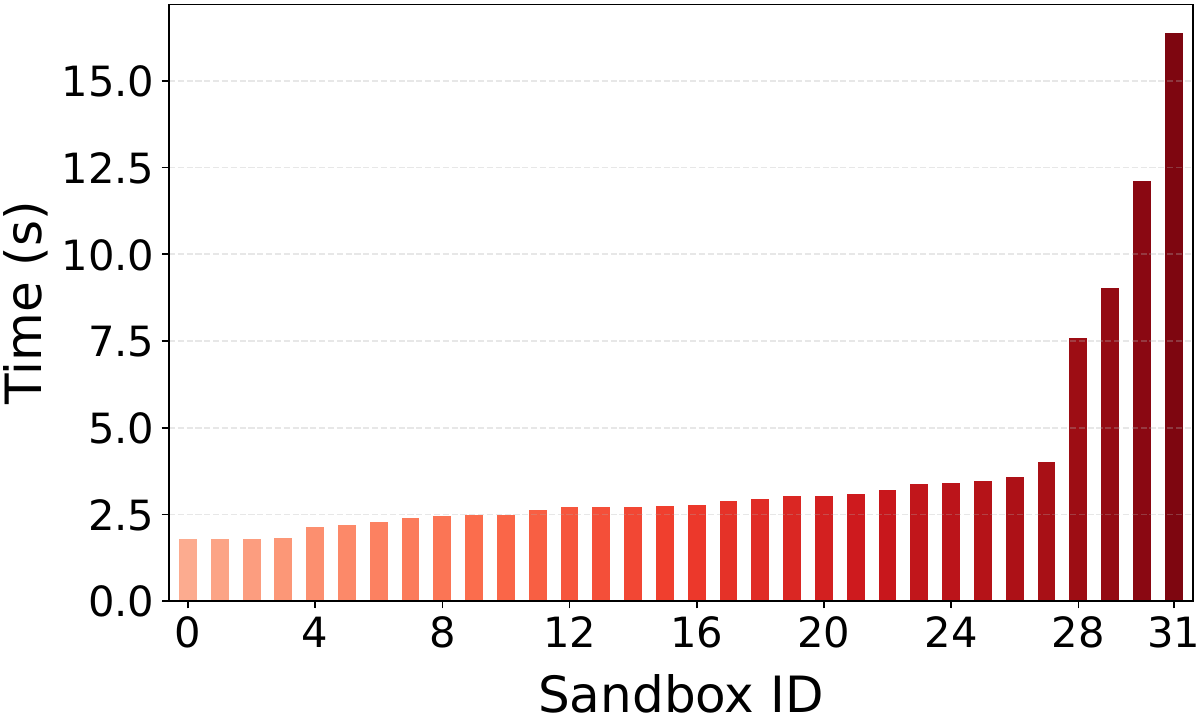}
        \caption{Sandbox cold-start latency.}
        \label{fig:cold_start_time}
    \end{subfigure}
    \vspace{-0.8em}
    \caption{Execution bottlenecks in agent runtimes. (a) Execution-step latency breakdown. (b) Mean cold-start latency across 32 sandboxed environments, including all MCPBench~\cite{wang2025mcp} sandboxes and additional commonly used environments. Most sandboxes initialize in 2--4 seconds, while resource-intensive environments can require up to approximately 20 seconds.}
    \vspace{-4mm}
    \label{fig:execution_bottlenecks}
\end{figure}

As illustrated in Fig.~\ref{fig:time_breakdown}, the end-to-end latency of a single execution step $N$ can be broken down into several parts:
\begin{equation}
T_{step}^{(N)} = T_{context}^{(N)} + T_{generation}^{(N)} + T_{env\_prep}^{(N)} + T_{data\_io}^{(N)} + T_{sandbox\_exec}^{(N)}
\end{equation}

The first two components are associated with LLM inference, including context processing and token generation. The remaining three define the optimization boundary for an agent runtime: $T_{env\_prep}^{(N)}$ is the time duration to get the selected sandbox environment ready, $T_{data\_io}^{(N)}$ is the time to exchange invocation inputs and results, and $T_{sandbox\_exec}^{(N)}$ is the time spent executing the tool. 

\SystemName targets these time frames without changing the LLM inference stack by overlapping preparation with reasoning, reducing data movement, and eliminating redundant executions. We derived the following observations that motivate this study.

\mypara{Observation \#1}: \textit{In-step preparation can overlap with token generation.}
In today's reactive execution model, environment initialization cannot start until the LLM emits a explicit sandbox invocation, even though the prompt, plan, and partial token stream often already reveal the likely sandbox. Using this partial evidence to start preparation earlier can put the prewarm even forward, ahead of the corresponding invocation. This interval lets sandbox startup, runtime connection setup, and other preparation work overlap with the remaining $T_{generation}^{(N)}$, rather than placing all of $T_{env\_prep}^{(N)}$ on the critical path. As shown in Fig.~\ref{fig:cold_start_time}, sandbox initialization requires several seconds and can approach 20 seconds for heavyweight sandboxes, rendering this overlap mechanism critical for minimizing the effective preparation latency observed by the system.

\mypara{Observation \#2}: \textit{Cross-step execution can advance the prewarming.}
Consecutive steps within an agent session frequently exhibit strong temporal locality. For instance, in a large language model (LLM) serving workload, a \texttt{paper} \texttt{search} step is often followed by a \texttt{document} \texttt{reading} step that processes a retrieved document, and a \texttt{data analysis} step is commonly followed by a \texttt{figure} \texttt{generation} step. Nonetheless, prevailing runtime systems generally treat each invocation as an independent event, discarding workflow context once a step is initiated. In contrast, the runtime can exploit the execution window of the current sandbox to proactively instantiate and prewarm the most likely subsequent execution environments before step $N+1$ begins, thereby making the intra-step prewarm more in-advance beyond the current decoding stage.

\mypara{Observation \#3}: \textit{Redundant execution and excessive unnecessary data transfer.}
In multi-turn agent workflows, later reasoning steps often revisit information that was already produced, such as querying the same document again or re-running deterministic analyses. However, conventional runtimes still re-execute these operations even when an equivalent result is already available, repeatedly paying the cost of $T_{sandbox\_exec}^{(N)}$. Meanwhile, sandbox interfaces often couple the synchronous transfer of large intermediate artifacts (e.g., files, images, structured outputs) with control messages, causing $T_{data\_io}^{(N)}$ to scale with artifact size. Intuitively, one can accelerate the agent execution by caching and reusing prior execution results and decoupling artifact transmission from control signaling, thereby eliminating redundant work and data transfer overhead.

These observations reveal optimization opportunities at three points in the agent execution loop: overlapping the environment preparation with the LLM operations within a step, advancing the environment preparation of the next step in the current step, and eliminating repeated execution or data movement. Existing LLM serving frameworks primarily focus on optimizing model execution --- such as prefill and decoding phases, as well as KV cache management --- while serverless runtimes primarily reduce the overhead associated with container initialization without considering user-wise intent. However, neither of them directly addresses how an agent-oriented runtime can leverage these optimizations while  maintaining resource efficiency and preserving the intended semantics of tool invocation and usage.

\begin{figure*}[t]
    \centering
    \includegraphics[width=0.8\textwidth]{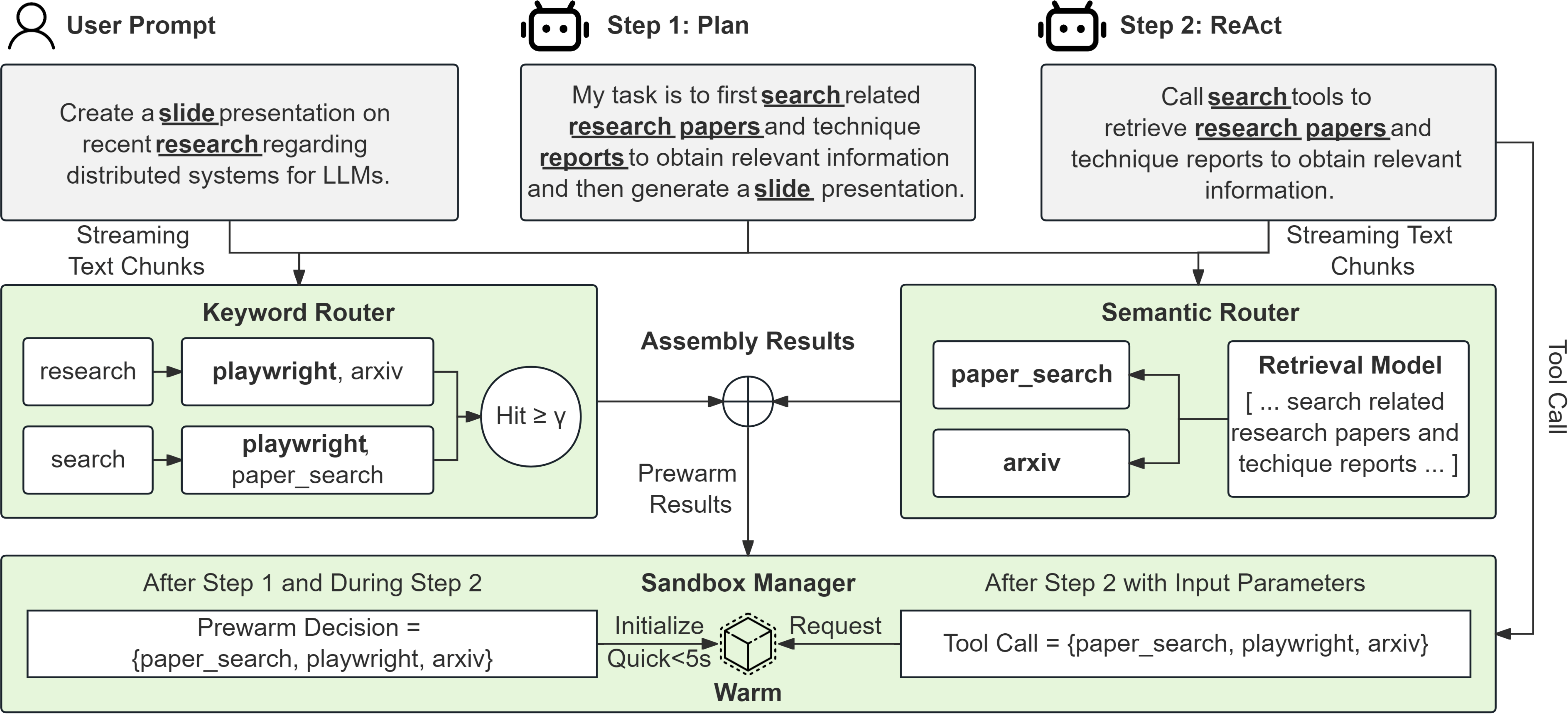}
    \vspace{-0.06em}
    \caption{Intent-aware sandbox prewarming: the plan agent expands user requests into ReAct steps, and keyword plus semantic predictions jointly decide to prewarm the sandbox or not.}
 \vspace{-4mm}
\label{fig:async_intent_routing}
\end{figure*}

\begin{figure}[t]
    \centering
\includegraphics[width=0.95\columnwidth]{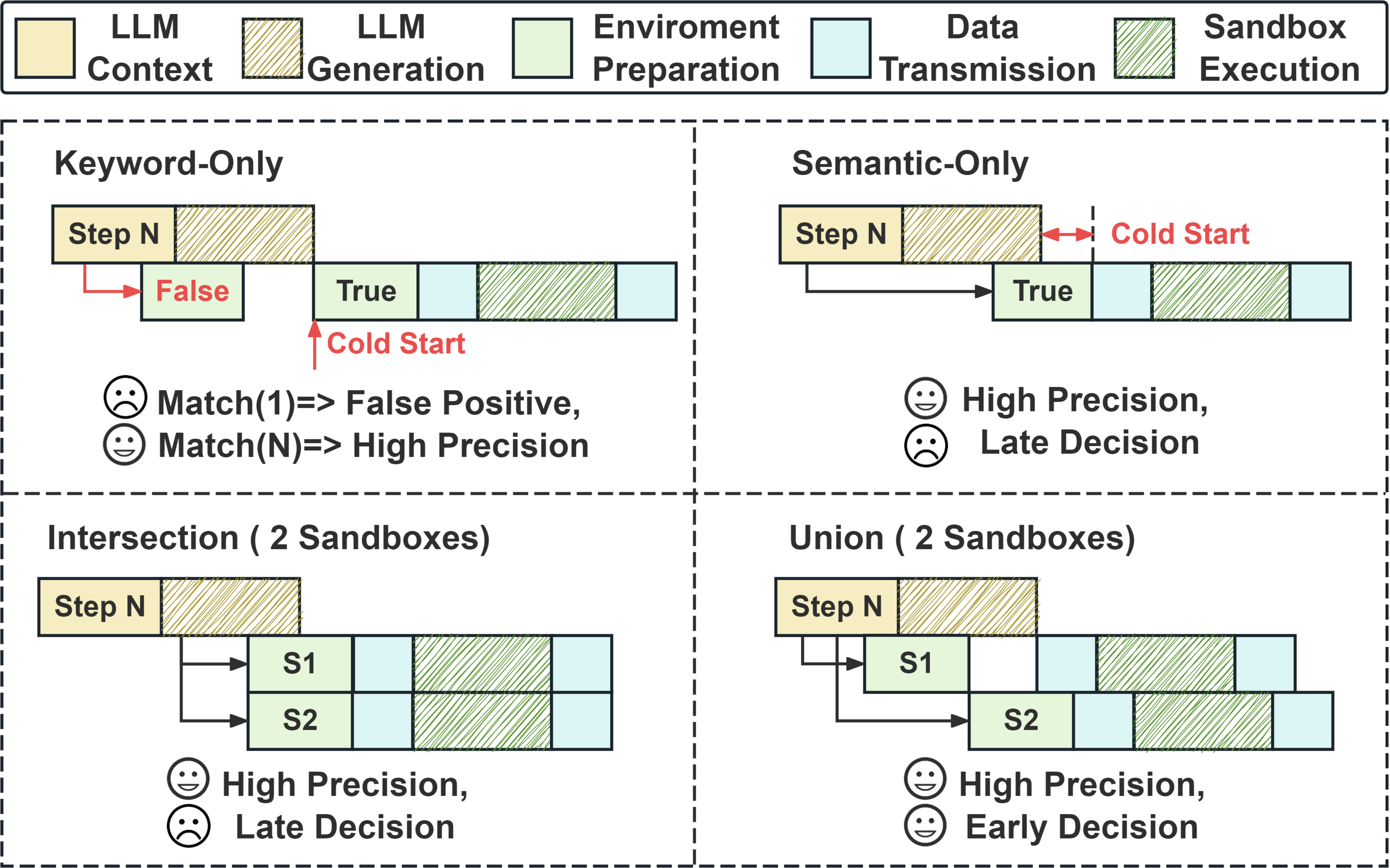}
    \vspace{-0.05em}
    \caption{Tradeoffs among routing policies. Keyword routing trades early decisions for potential false positives; semantic routing and intersection delay a precise decision; union assembly retains precision while allowing either router to trigger prewarming early.}
 \vspace{-4mm}
\label{fig:routing_tradeoff}
\end{figure}

\vspace{-0.2em} 
\subsection{Research Challenges}
Designing an effective execution runtime to accelerate the LLM agent serving is faced with the following challenges. 


\mypara{Challenge \#1}: \textit{Early intent prediction vs. resource waste.} Intent speculation of tool use from given incomplete token stream is critical to ensure the timeliness of sandbox environment preparation whilst avoiding unwanted tool launch due to inaccurate intent prediction. Early warmup must rely on highly ambiguous early-stage tokens (e.g., general-purpose verbs such as \texttt{read} and \texttt{get} reused across different MCP tools), which often causes \textit{false-positive} sandbox activation, wasting resources and potentially triggering host-level OOM in multi-tenant deployments.

\mypara{Challenge \#2}: \textit{Cross-step in-advance prewarming vs. non-deterministic workflow execution.}  It is advantageous to select the subsequent-step sandbox environments in advance of the manifestation of the next intent. The agent workflow executes in a probabilistic manner, admitting multiple plausible successor states. Prewarming all potential successors incurs a computational and resource overhead comparable to that of a fully reserved deployment, whereas prewarming only a single successor substantially reduces the opportunity to amortize sandbox initialization costs and yields only limited latency improvements.  Instead, the runtime system should exploit historical transition data to rank a small set of highly probable target environments, then refine this choice using the signal available at the current step. This balances resource utilization against latency reduction while preserving the ability to conceal sandbox initialization behind ongoing computation.

\mypara{Challenge \#3}: \textit{Reuse and data-path efficiency vs. sandbox compatibility.}
We must skip redundant work without changing what the agent observes from a sandbox invocation. Exact reuse is safe but misses semantically equivalent requests expressed differently; unconstrained semantic reuse may yield incompatible results. Similarly, putting large artifacts in normal RPC messages preserves compatibility but keeps data movement proportional to payload size, while replacing the control interface would break existing tools. Hence, the runtime must detect compatible reuse, reject unsafe approximations, and decouple bulky artifacts from control signaling while preserving sandbox-execution semantics.

\section{Design of \SystemName}
\label{sec:ourapproach}

We focus on three critical time periods within the execution of reactive agents: environment preparation, data movement, and sandbox execution. This section how \SystemName accelerate them with decoupled yet inter-connected optimizations: intent-aware sandbox prewarming that infers an in-step tool intent early to overlap the environment preparation with the decoding of LLM serving (\S~\ref{sec:appr:intent_routing}); stochastic sandbox prefetching that exploits cross-step workflow regularity to further put the preparation forward (\S~\ref{sec:appr:sandbox_prefetching}); and reuse-aware data transmission that avoids redundant execution and removes large artifacts from the control path (\S~\ref{sec:appr:reuse_aware_transmission}).

\begin{figure*}[t]
    \centering
    \includegraphics[width=0.8\textwidth]{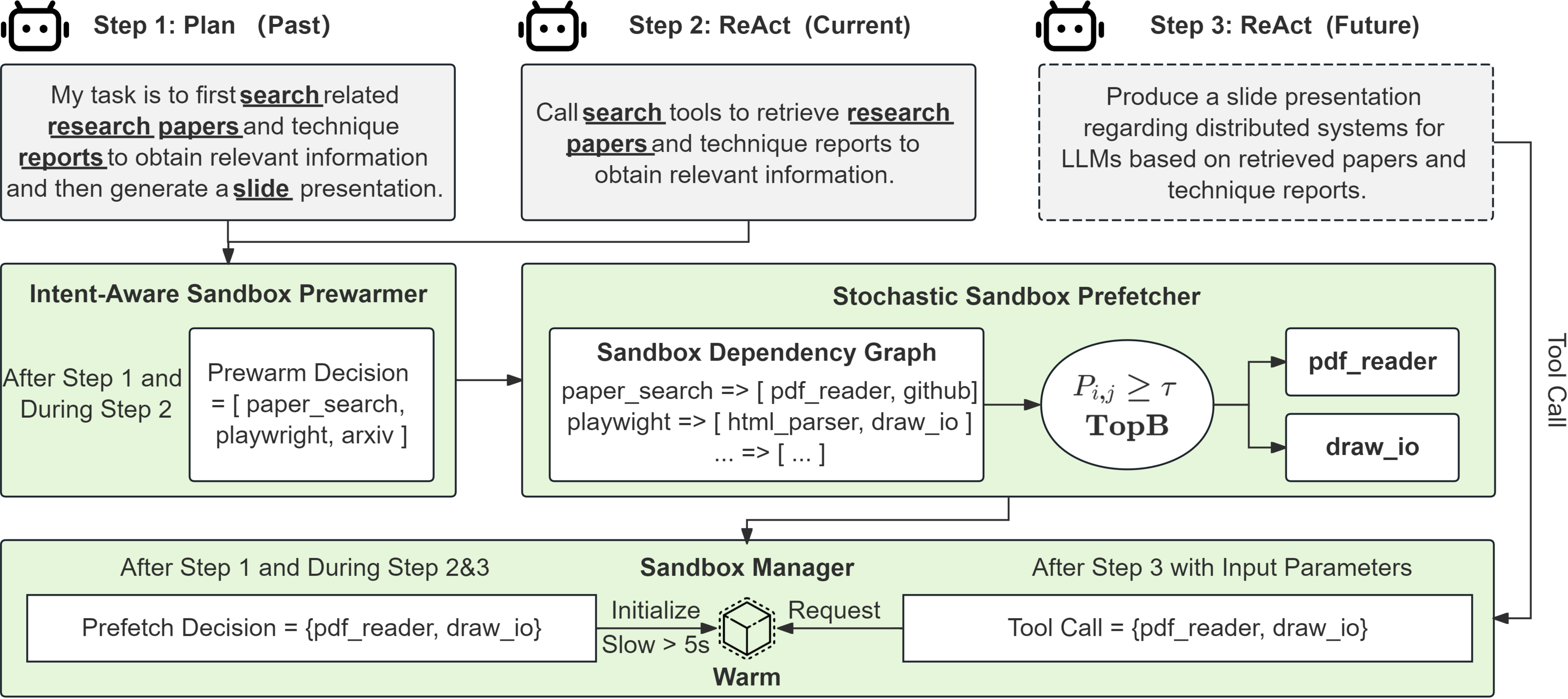}
    \vspace{-0.6em}
    \caption{Stochastic sandbox prefetching in \SystemName. Execution history predicts likely sandboxes for subsequent ReAct steps.}
\label{fig:stochastic_sandbox_prefetching}
\end{figure*}

\subsection{Intent-Aware Sandbox Prewarming}
\label{sec:appr:intent_routing}

Naive reactive runtimes await a full tool invocation before sandbox preparation, even if prompts, evolving plan, and partial generation already reveal user intents. \SystemName instead prewarms the sandbox earlier by inferring streaming token-level intents, better overlapping sandbox preparation with ongoing LLM inference.

\SystemName makes the best use of the LLM docoding period $T_{generation}^{(N)}$  to obtain \textit{just enough} user intents for launching accurate sandboxes without compromising the timeliness of prewarming. This is done by streaming context to two independent routers. As shown in Fig.~\ref{fig:routing_tradeoff}, there exists a dilemma: earlier intent predictions overlap more with the LLM decoding period but are less reliable and can waste sandbox capacity. Keyword Router acts early but must balance false positives against a stricter threshold; Semantic Router is more precise but slower. Requiring both predictions incurs the Semantic Router's delay, whereas opportunistically combining their outputs preserves high precision while allowing earlier decisions. Hence, \SystemName proposes an intent-aware sandbox prewarming mechanism that combines the two asynchronous intent predictions, rather than considering either one on its own to be sufficient.

\cparagraph{Keyword Router}
The Keyword Router scans the stream against tool-specific keyword profiles and can emit a candidate within microseconds of a distinctive token, preparing the environment while the LLM continues decoding. Common terms like \texttt{research}, \texttt{search}, or \texttt{slide} occur in many tool descriptions: triggering on a single match creates a large prewarm set with many false positives, while requiring many matches will delay the preparation, missing good chances of overlapping. We therefore apply a threshold $\gamma$ on the number of matched tool-specific keywords. In Fig.~\ref{fig:async_intent_routing}, surface cues in the user request and evolving plan let the Keyword Router emit likely sandbox candidates before the plan agent finishes the current ReAct step. The threshold balances early activation with resource waste; the chosen configuration is given in \S~\ref{sec:exp:ablation}.

\cparagraph{Semantic Router}
In parallel, the Semantic Router compares the active context with tool-intent representations. It captures requests whose wording overlaps little with a tool profile and disambiguates generic keyword cues using the plan and prior generation. In Fig.~\ref{fig:async_intent_routing}, it can identify candidate sandboxes from the broader task intent even when no single token uniquely identifies a tool. Its tradeoff is temporal: reliable semantic evidence typically needs a longer prefix, so a semantic-only decision often arrives too late to hide the cold start latency of a heavy-weight sandbox such as PaperSearch~\cite{papersearch}, and Neo4j~\cite{neo4j}. The Semantic Router is an asynchronous, complementary source of candidates that can recover intents the Keyword Router misses.

\cparagraph{Union Assembly}
Let $\mathcal{S}_{key}^{(N)}$ and $\mathcal{S}_{semantic}^{(N)}$ be the candidate sets from the two routers at token step $N$. The lower-left quadrant of Fig.~\ref{fig:routing_tradeoff} shows why prewarming only their intersection is unsuitable: it boosts apparent precision but makes every trigger wait for the Semantic Router and drops valid tools whenever either router has imperfect recall.  Instead, we use the lower-right policy:
\begin{equation}
\mathcal{S}_{trigger}^{(N)} = \mathcal{S}_{key}^{(N)} \cup \mathcal{S}_{semantic}^{(N)}.
\end{equation}
Each router manages its own false positives, and the union lets the first credible signal start preparation. Fig.~\ref{fig:async_intent_routing} shows the resulting behavior: routers independently generate candidates from the request and partial plan, then form a unified prewarm set for the sandbox manager. Explicit intents benefit from the early prewarm of the Keyword Router, while implicit intents are handled by the Semantic Router. As will be shown in \S~\ref{sec:exp:ablation}, asynchronous union can minimize waiting time and eliminate cold starts in the evaluated routing workload.

\begin{figure}[t]
\centering
\includegraphics[width=1.05\columnwidth]{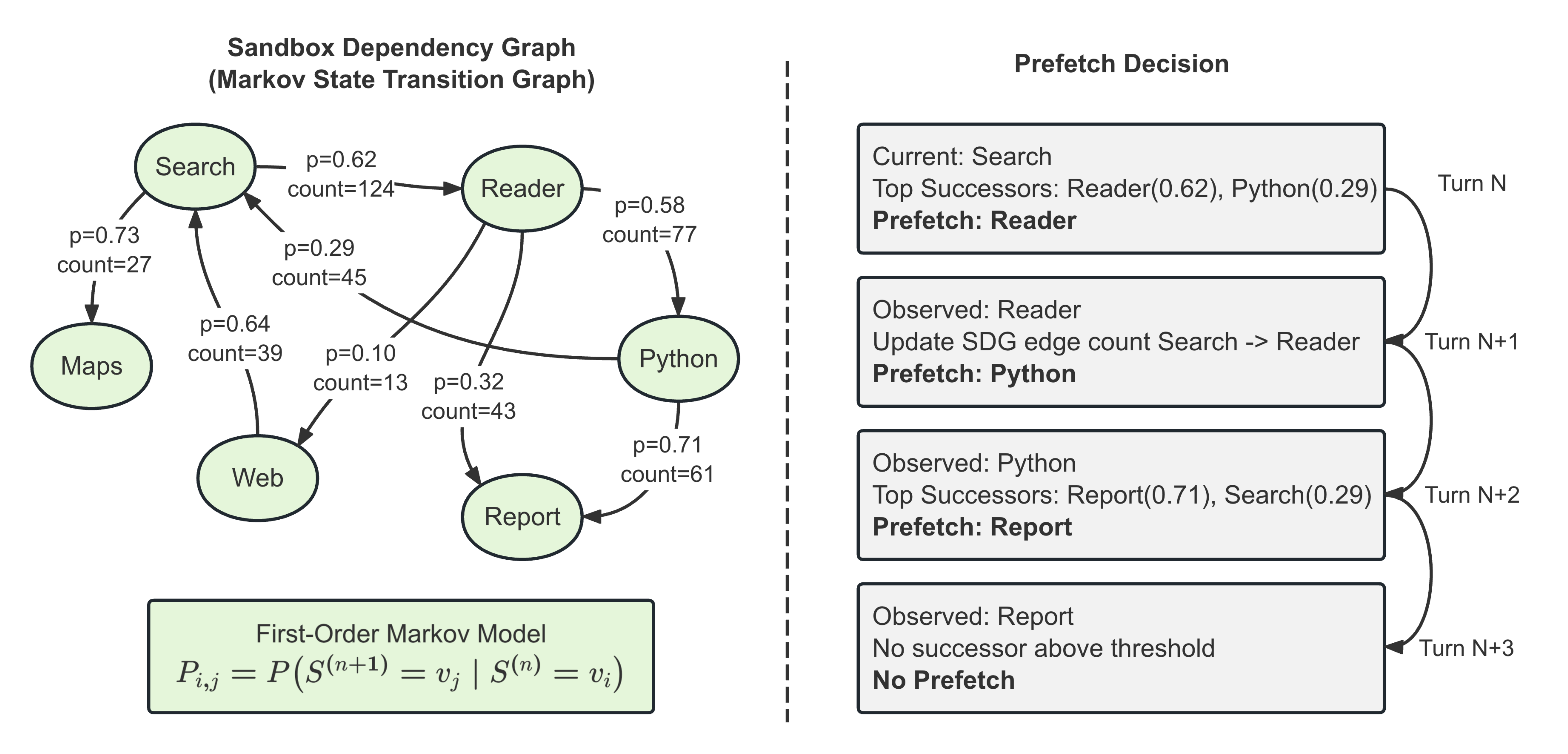}
 \vspace{-0.4em}
 \caption{Stochastic sandbox prefetching using a first-order Markov model. Left: SDG-based Markov state transition graph with edge probabilities from observed counts. Right: example session showing thresholded, budgeted prefetching across turns, where predicted successors are prewarmed before the next step commits.}
\vspace{-4mm}
\label{fig:markov_prefetch_example}
\end{figure}

\subsection{Stochastic Sandbox Prefetching}
\label{sec:appr:sandbox_prefetching}

Intent-aware sandbox prewarming can only use the remaining decoding time of the current step, which is inadequate when a non-resident sandbox takes longer to start than the marginal token-generation window. In reality, steps in an agent workflow are non-deterministic but generally follow a probabilistic model. For instance, after a paper search, an agent may read a returned document; after data analysis, it may generate a figure or report. \SystemName exploits these cross-step probabilistic patterns to navigate the environment preparation in advance during $T_{sandbox\_exec}^{(N)}$, before the next step commits to a specific invocation.
\cparagraph{Stochastic Markov Process Modeling} Let $\mathcal{V}$ denote the sandbox state space, where each state is a sandbox type or a typed tool-state tuple in the state-level SDG. 
For an execution trace $\{S^{(1)}, \dots, S^{(N)}\}$, \SystemName records step-to-step transitions in a directed sandbox dependency graph (SDG).
For each ordered pair $(v_i, v_j)$, we maintain transition counts:
\begin{equation}
C_{i,j} \leftarrow C_{i,j} + \mathbf{1}\!\left[S^{(n)}=v_i,\; S^{(n+1)}=v_j\right]
\end{equation}
The next-state probability is estimated by a first-order Markov model with \textit{Laplace Smoothing}~\cite{laplace_thorie_1812}:
\begin{equation}
P_{i,j}
= P\!\left(S^{(n+1)}=v_j \mid S^{(n)}=v_i\right)
= \frac{C_{i,j}+\alpha}{\sum_{k \in \mathcal{V}} (C_{i,k}+\alpha)},
\end{equation}
where $\alpha=1$ avoids zero-probability collapse for sparsely observed transitions.

\cparagraph{Prefetching Under Budget Constraints} Given current state $v_i$, \SystemName first identifies sandbox candidates with non-trivial cold-start cost:
\begin{equation}
\mathcal{K}_i=\left\{v_j \in \mathcal{V} \setminus \{v_i\}\;\middle|\;L_j \ge \lambda\right\},
\end{equation}
where $L_j$ denotes the estimated cold-start penalty of sandbox $v_j$ and $\lambda$ is a lightweight cost threshold.

\SystemName then filters out low-confidence successors and retains only high-probability candidates:
\begin{equation}
\mathcal{C}_i=\left\{v_j \in \mathcal{K}_i\;\middle|\;P_{i,j} \ge \tau\right\},
\end{equation}
where $\tau$ controls false-positive prewarming.
Finally, \SystemName selects the top-$B$ sandboxes under a fixed budget:
\begin{equation}
\mathcal{A}_i=\operatorname{top-B}\!\left(\mathcal{C}_i,\, P_{i,j}\right),
\end{equation}
where $\mathcal{C}_i$ is the filtered candidate set, $\mathcal{A}_i$ is the final budgeted prefetch set, and $B$ bounds the number of sandboxes prewarmed per step. This policy is intentionally lightweight and can be executed outside the LLM generation critical path.

\cparagraph{Online Update}
After each completed sandbox invocation, \SystemName appends one transition edge to SDG and updates $C_{i,j}$ and $P_{i,j}$ asynchronously in the background prefetch worker. Thus, the predictor continuously adapts to evolving multi-turn workflows without blocking foreground agent execution. Fig.~\ref{fig:stochastic_sandbox_prefetching} shows how stochastic sandbox prefetching operates together with intent-aware sandbox prewarming in a complete multi-step workflow. After step $N$ finishes in sandbox $v_i$, \SystemName queries the SDG for likely successor sandboxes, filters out low-value or low-confidence candidates using $L_j \ge \lambda$ and $P_{i,j} \ge \tau$, and then selects a budgeted prefetch set $\mathcal{A}_i$ before step $N+1$ is fully committed.

\begin{figure}[t]
    \centering
    \includegraphics[width=0.9\columnwidth]{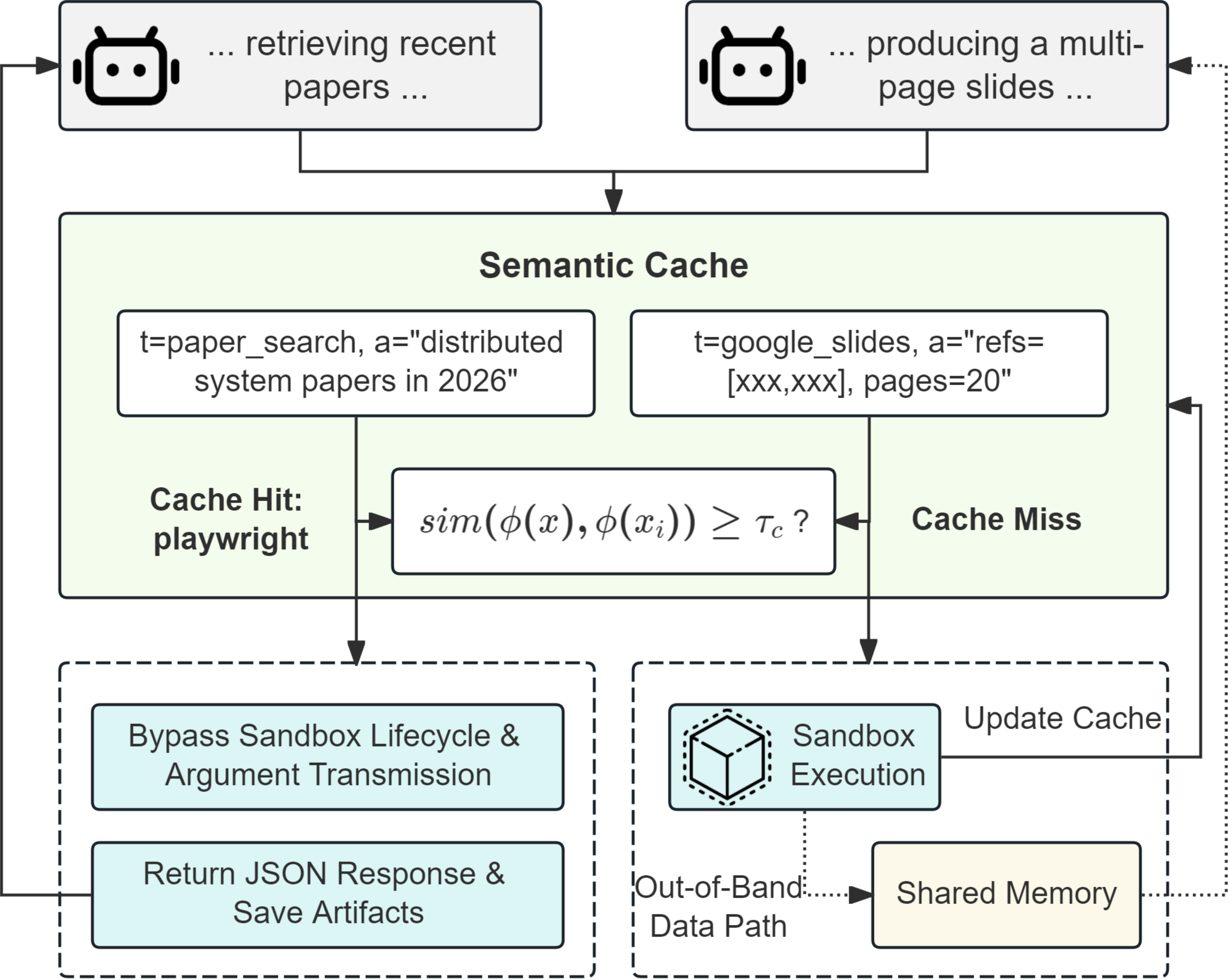}
    \caption{Reuse-aware data transmission in \SystemName. A semantic-cache hit returns a prior result \text{(paper\_search)}; a miss executes the sandbox \text{(paper\_slides)} and delivers the result through the out-of-band data path.}
 \vspace{-4mm}
\label{fig:data_plane_optimization}
\end{figure}

\subsection{Reuse-Aware Data Transmission}
\label{sec:appr:reuse_aware_transmission}

Once an invocation is ready, its remaining critical-path cost lies in artifact movement ($T_{data\_io}^{(N)}$) and tool execution ($T_{sandbox\_exec}^{(N)}$). \SystemName first checks if it can reuse a cached result; otherwise, it runs the tool and returns its artifact via a separate data path. Fig.~\ref{fig:data_plane_optimization} details this procedure.

\cparagraph{Semantic Caching} Multi-turn agents often re-access unchanged documents, submit semantically equivalent queries with different wording, or re-run deterministic computations. Exact argument matching is robust but fails to capture such superficial variations, while unconstrained semantic matching can mistakenly merge requests that differ in tools, inputs, or side effects. Thus, caching must broaden the set of reusable requests while strictly preserving functional equivalence and behavioral compatibility.

For each completed deterministic invocation, \SystemName stores a normalized invocation signature and its result. During lookup, it first filters cache entries by tool identity, then compares normalized invocation representations are compared against cached signatures. For a request $x$, result reuse is allowed only if a compatible cache entry exists that satisfies the semantic equivalence constraint:
\begin{equation}
\mathrm{hit}(x) = \exists i \;\text{s.t.}\; tool(x)=tool(x_i) \;\wedge\; sim(\phi(x),\phi(x_i)) \geq \tau_c,
\end{equation}
where $x$ denotes the incoming tool invocation request and $x_i$ denotes the $i$-th cached invocation. Function $\phi(\cdot)$ transforms an invocation into a normalized semantic representation by first removing superficial variations in argument representation and then extracting semantic features for similarity comparison. Function $sim(\cdot,\cdot)$ measures the similarity between two invocation representations, and the threshold $\tau_c$ controls the strictness of semantic reuse. The tool identity constraint guarantees interface-level compatibility, while the similarity threshold limits reuse to invocations with sufficiently similar normalized semantics.

This design treats semantic matching as a conservative extension of exact reuse rather than an unconstrained approximation mechanism. Semantic similarity only enlarges the reusable request space within the boundary of the same tool interface and deterministic invocation behavior. On a cache miss, execution proceeds along the standard path and appends the fully computed result to the cache. On a cache hit, both sandbox initialization and tool invocation are skipped, thereby reducing $T_{env\_prep}^{(N)}$ and $T_{sandbox\_exec}^{(N)}$. As will be demonstrated in \S~\ref{sec:exp:semantic_cache}, the semantic caching strategy can recover a larger proportion of redundant computation than using the strategy of exact matching alone, while maintaining a conservative fallback path that preserves the inference correctness.

\cparagraph{Out-of-Band Data Transmission}
Even after a sandbox is fully initialized, conventional RPC transports serialize large logs, files, images, and structured outputs into request–response messages. Consequently, $T_{data\_io}^{(N)}$ scales with the payload size and blocks the agent prior to its subsequent reasoning step. A wholesale replacement of the control protocol would compromise compatibility with existing MCP tools; therefore, \SystemName instead decouples control signaling from artifact transfer.

The in-band control plane continues to carry standard tool directives, completion notifications, error reports, and compact metadata. Large artifacts are represented on this plane solely by a fixed-size reference and are exchanged via a co-located, zero-copy data path. Under this design, a cache miss executes as usual, publishes its artifact exactly once, and returns a reference to the Agent Engine; a cache hit simply reuses and returns the previously published artifact over the same data path. This architectural decoupling maintains the existing control interface while eliminating RPC-layer serialization and memory copying for high-volume outputs. As demonstrated in \S~\ref{sec:exp:oob}, the corresponding transmission latency becomes effectively insensitive to payload size.
\section{Implementation}
\label{sec:implementation}

\begin{figure}[t]
  \centering
  \includegraphics[width=0.95\columnwidth]{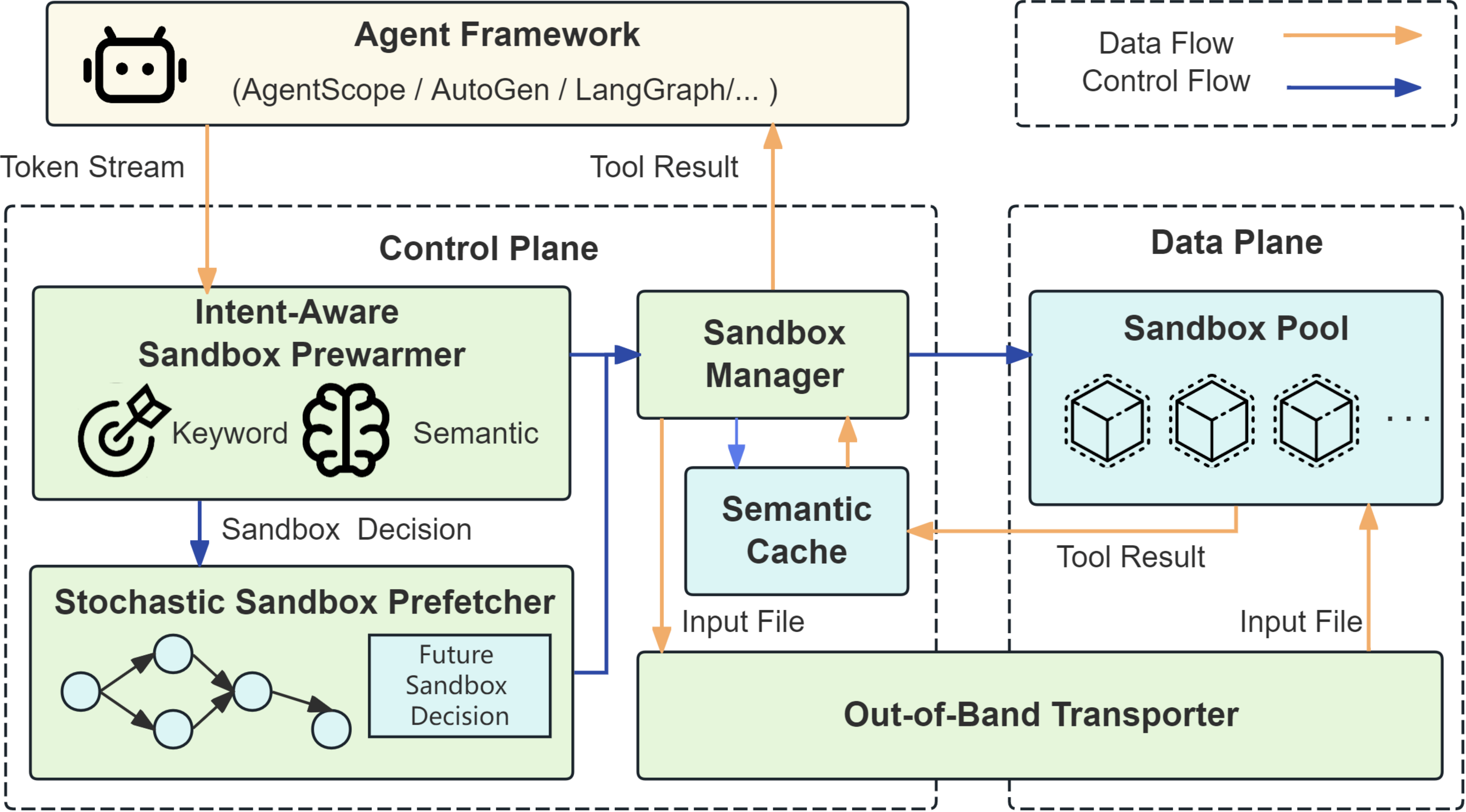}
  \vspace{-0.3em}
  \caption{Overview of \SystemName.}
\label{fig:overview}
\end{figure}

\SystemName is implemented as a predictive serving runtime between the agent engine and external execution sandboxes. Figure~\ref{fig:overview} shows its control and data paths. The implementation observes an agent run-loop without changing its tool-facing interface, and overlaps runtime work with the LLM and sandbox work already in progress.

\cparagraph{Control Plane}
On the control plane, a controller subscribes to the incoming token stream and records completed tool transitions. The Keyword Router matches the stream against tool keyword profiles and emits a candidate when the number of matched tool-specific keywords reaches $\gamma=2$, the operating point selected in Section~\ref{sec:exp:ablation}. In parallel, the Semantic Router uses the sparse retrieval configuration selected in that section. Union Assembly dispatches their candidate sets independently to the sandbox manager, which deduplicates requests and starts the corresponding containers. A background prefetch worker updates the sandbox dependency graph from execution traces, applies the transition probabilities in Section~\ref{sec:appr:sandbox_prefetching}, and submits budgeted next-step warmups while the current tool is executing. In our implementation, we configure the cost threshold as $\lambda=5$. prefetch probability threshold as $\tau=0.6$ and the per-step prefetch budget as $B=1$.

\cparagraph{Data Path}
Before scheduling a deterministic invocation, the runtime constructs its invocation representation, restricts lookup to the same tool identity, and checks its semantic-result cache. Each cache entry contains the tool identifier, normalized invocation signature, semantic embedding, and result reference. The cache reuses a previous result only when the tool identity matches and the semantic similarity exceeds the threshold $\tau_c=0.8$ used in the reported experiments. On a miss, the sandbox writes a large result into a host-managed memory-mapped shared-memory region. The control plane carries only a 64-bit \texttt{token\_id} that names this region, while the Agent Engine reads the result directly from the shared-memory backplane. Small directives, metadata, completion notifications, and errors remain on the normal control plane. This implements the common cache-or-execute flow in Figure~\ref{fig:data_plane_optimization} without serializing bulky results through the RPC stack.

\cparagraph{Execution boundary and correctness}
We consider multi-tenant agents whose tools run in OS-isolated containers or microVMs. Predictive preparation makes an environment ready but does not execute side-effecting work before the agent commits the invocation; unused warmups are discarded. Cache reuse is limited to deterministic, compatible tool requests. The shared-memory bridge operates inside one trusted host or securely managed cluster boundary, with existing access controls preventing cross-tenant memory access. Sandbox escapes, malicious agents, and compromised infrastructure are outside this work's threat model.
\vspace{-0.2em}
\section{Evaluation}
\label{evaluation}

\subsection{Experiment Setup}
\label{sec:exp:setup}
\cparagraph{Hardware and Software Environment}
All experiments are conducted on a commodity server equipped with an 16-core CPU, 256\,GiB of host memory, and a 2\,TB NVMe SSD. \SystemName is built upon the AgentScope~\cite{gao2024agentscope} framework, specifically integrated with its agent engine layer. The isolated multi-tenant execution sandboxes are instantiated via Docker containers, and the entire runtime infrastructure is implemented in Python. To drive agent reasoning, we utilize Alibaba DashScope's Qwen3.5-Max model~\cite{dashscope2026}, accessed concurrently via its production cloud API endpoints.

\cparagraph{Workloads}
We evaluate \SystemName using a trace-level benchmark with 200 multi-turn trajectories, derived from MCPBench~\cite{wang2025mcp} with 32 open-source MCP-compatible tool servers (e.g., Playwright~\cite{playwright}, Jupyter~\cite{jupyter}, Neo4j~\cite{neo4j}) collected from GitHub~\cite{githubmcp}. To ensure tool-level validity and realistic workflow execution dependencies, we construct the benchmark in two stages: 
\begin{itemize}[leftmargin=1.2em]
    \item \textbf{Atomic Tool-use Generation:} We generate 20 single-turn interaction templates per tool (640 in total) to cover diverse atomic behaviors across all 32 tools.
    \item \textbf{Multi-turn Trajectory Construction:} An LLM planner incrementally generates 1--10 step agent sessions, where each planned task is executed against actual servers to get real execution results back into subsequent planning steps.
\end{itemize}
The resulting dataset is uniformly distributed with 20 traces per trajectory length, averaging 6.4 steps per session and 2.96 tools per step. This execution-grounded approach ensures tool-level validity and eliminates tool-specific bias while reflecting representative LLM agent workflows.

\cparagraph{Methodology}
Unless otherwise specified, we evaluate all systems by replaying full traces as session-level workloads with a fixed random seed (\texttt{seed} = 0). Each trace is executed in a step-wise manner, preserving dependencies across reasoning, tool invocation, and sandbox execution to faithfully model realistic multi-turn agent workflows. Under this deterministic sampling setting, the resulting trace distribution naturally exhibits a concentration in the 5--8 step range, which we therefore treat as the representative workload regime rather than a manually selected subset.

We evaluate all experiments at the session level unless explicitly stated otherwise, and defer workload-specific configurations (e.g., trace length or sampling range) to each individual ablation study.

\begin{figure}[t]
\centering
\includegraphics[width=0.7\linewidth]{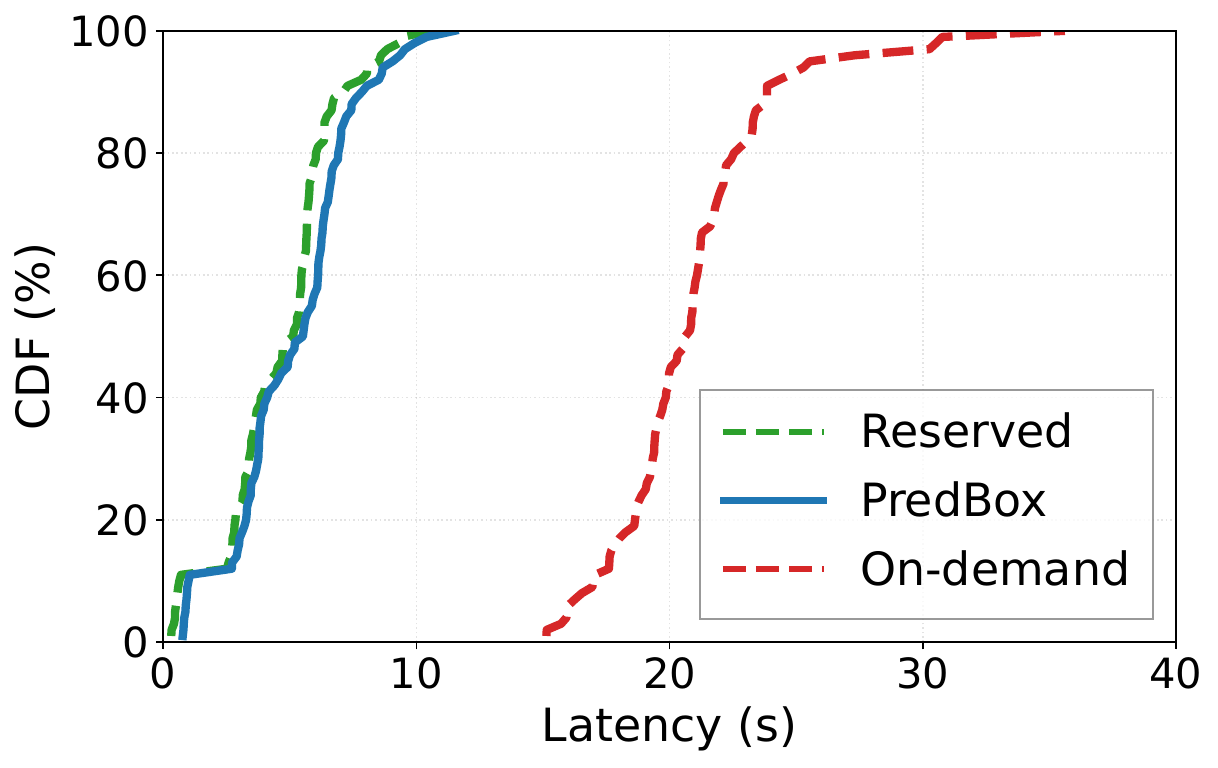}
\vspace{-0.3em}
\caption{Cumulative sandbox provisioning latency across multi-turn agent sessions (5--8 steps per session).}
\label{fig:cumulative_latency}
\end{figure}

\begin{figure*}[t]
\centering
\begin{subfigure}{0.245\linewidth}
    \centering
    \includegraphics[width=\linewidth]{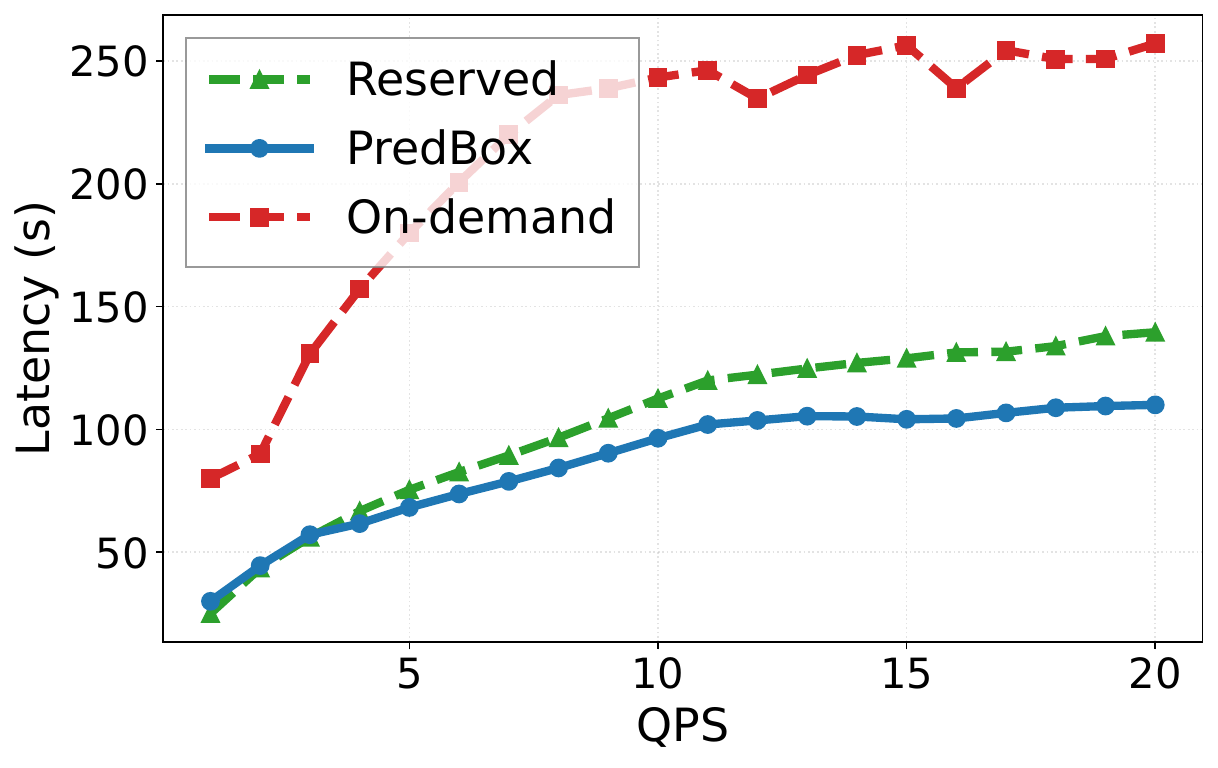}
    \caption{P99 E2E Latency}
    \label{fig:e2e_scalability:latency}
\end{subfigure}
\hfill
\begin{subfigure}{0.245\linewidth}
    \centering
    \includegraphics[width=\linewidth]{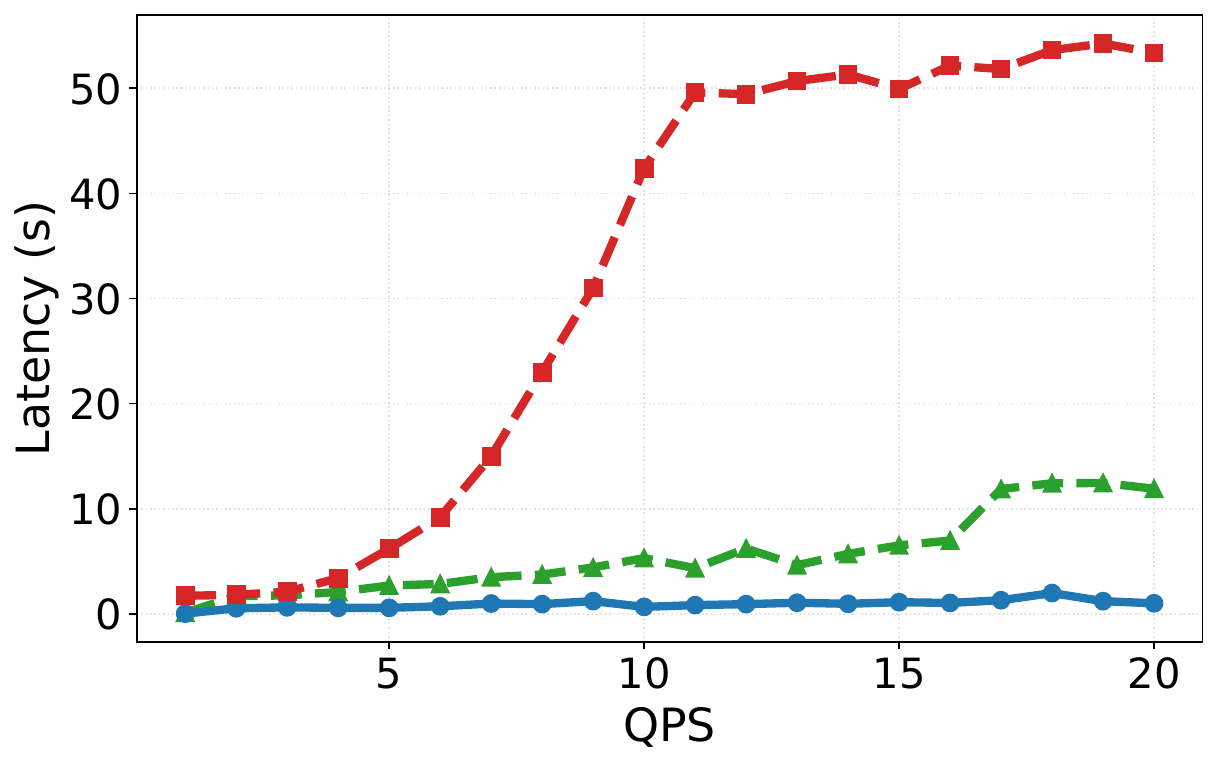}
    \caption{Cumulative Provisioning Latency}
    \label{fig:e2e_scalability:cumulative}
\end{subfigure}
\hfill
\begin{subfigure}{0.245\linewidth}
    \centering
    \includegraphics[width=\linewidth]{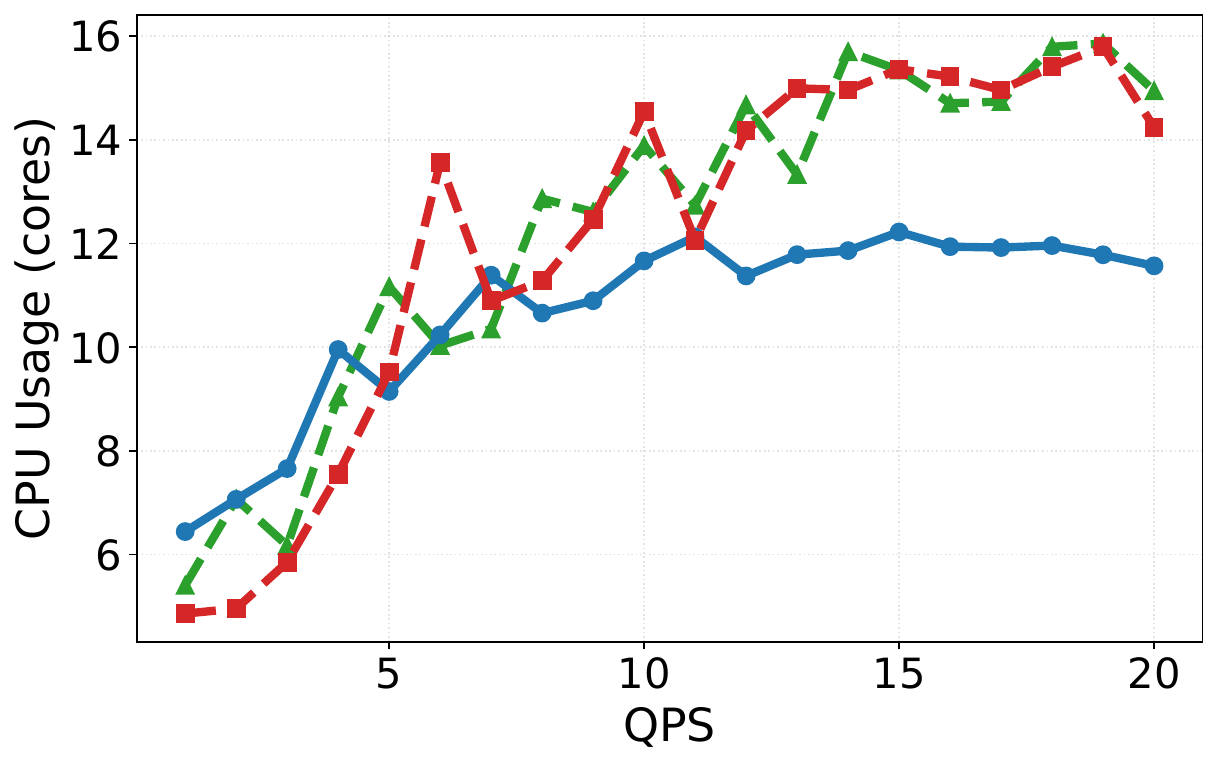}
    \caption{Peak CPU Usage}
    \label{fig:e2e_resource:cpu}
\end{subfigure}
\hfill
\begin{subfigure}{0.245\linewidth}
    \centering
    \includegraphics[width=\linewidth]{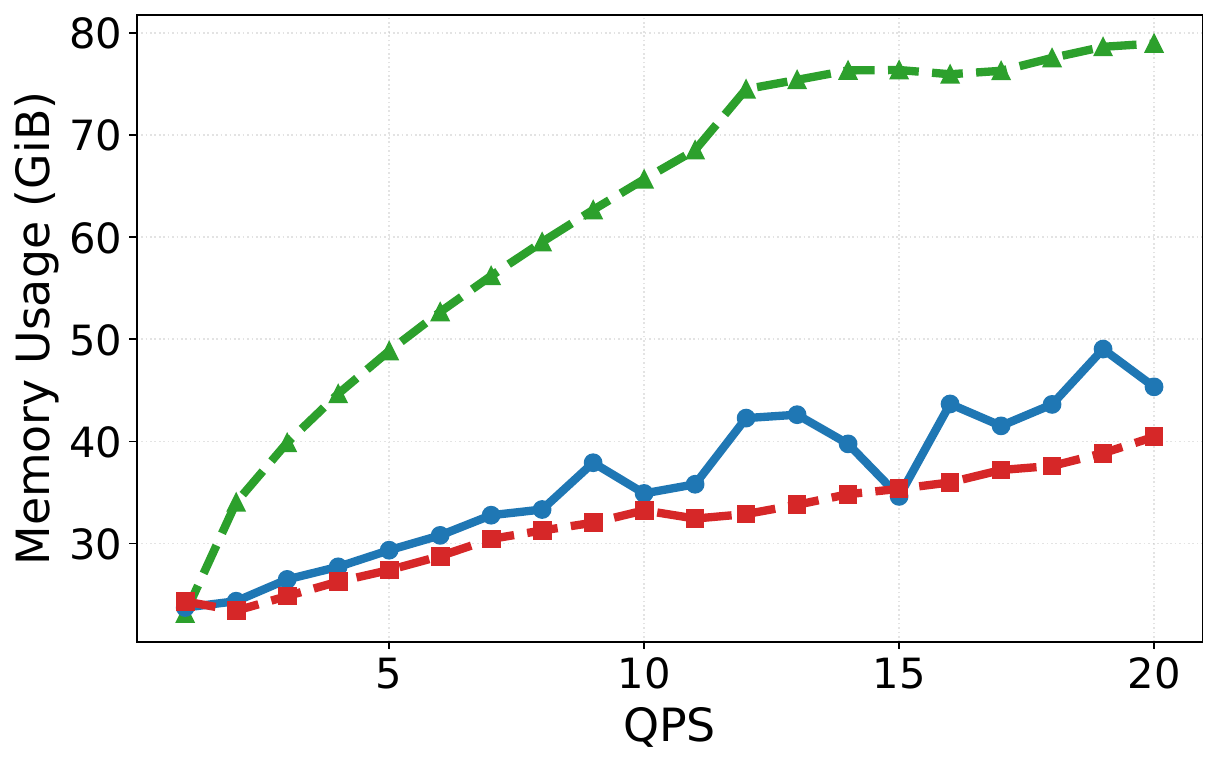}
    \caption{Peak Memory Usage}
    \label{fig:e2e_resource:memory}
\end{subfigure}
\vspace{-0.3em}
\caption{
End-to-end performance and resource consumption under concurrent workloads:
(a) P99 E2E latency,
(b) mean cumulative sandbox provisioning latency,
(c) peak CPU usage, and
(d) peak memory usage.
}
\label{fig:e2e_scalability_resource}
\vspace{-4mm}
\end{figure*}

\cparagraph{Metrics}
We evaluate \SystemName across three dimensions aligned with system design goals:
\begin{itemize}[leftmargin=1.2em]
    \item \textbf{End-to-end Latency:} Per agent session, we report mean and tail latency (P99) to capture both average performance and long-tail behavior under multi-step execution.
    \item \textbf{Resource Efficiency:} We measure peak CPU and memory consumption under multi-tenant workloads to quantify the runtime overhead introduced by sandbox provisioning, execution, and lifecycle management.
   \item \textbf{Prediction Accuracy:} We evaluate the accuracy of all predictive runtime optimizations, including intent-aware sandbox prewarming, stochastic sandbox prefetching, and semantic cache. For each, we report i) correct triggering or retrieval decisions under uncertain tool intent, and ii) false positive rate, reflecting unnecessary or incorrect activations such as misrouted intents, redundant prewarming.
\end{itemize}

\subsection{End-to-End Performance}
\label{sec:exp:end_to_end}

This section evaluates the end-to-end (E2E) performance of \SystemName under two representative baselines in agent serving. We define our baselines as follows:
\begin{itemize}[leftmargin=1.2em]
    \item \textbf{Reserved Runtime:} Maintains permanently warm sandboxes for all candidate tools. This represents the latency lower bound (performance ceiling). However, it is financially and physically non-viable in production due to prohibitive idle memory footprint across massive, multi-tenant MCP tool ecosystems.
    \item \textbf{On-demand Runtime:} Instantiates sandboxes dynamically upon tool calls, mirroring production MCP runtimes~\cite{aws2026,azure2026,gcp2026}. Since its bottleneck stems from application-level tool binding and session handshakes rather than generic OS booting, it remains the standard practical baseline.
\end{itemize}

We aim to answer a fundamental question: \textit{Can \SystemName successfully decouple execution latency from physical resource constraints, achieving serverful-like performance with serverless-like cost?}

\subsubsection{Cumulative Sandbox Provisioning Latency in Multi-Turn Agent Workflows}
\label{sec:exp:cumulative_latency}

We evaluate the end-to-end impact of \SystemName on sandbox provisioning latency over multi-turn agent execution traces. We define cumulative sandbox provisioning latency as the sum of sandbox initialization delays incurred at each tool invocation within a session, focusing exclusively on environment setup overheads and excluding in-sandbox computation time.

Fig.~\ref{fig:cumulative_latency} shows the distribution of cumulative latency across sessions. The \textit{On-demand} baseline exhibits steadily increasing latency over longer execution horizons due to repeated cold-start overheads, resulting in a heavy-tailed distribution. In contrast, \SystemName significantly reduces cumulative latency and maintains a much tighter distribution concentrated in the sub-second range. Compared to the \textit{On-demand} baseline, \SystemName achieves a $4.53\times$ reduction in cumulative sandbox provisioning latency. Against the \textit{Reserved} baseline, \SystemName remains within a $10.6\%$ performance gap while avoiding the substantial resource overhead of persistent sandbox allocation. These results demonstrate that predictive orchestration can effectively approximate \textit{Reserved} execution performance under a \textit{On-demand} deployment model.


\subsubsection{Scalability}
\label{sec:exp:e2e_scalability}

Fig.~\ref{fig:e2e_scalability:latency} summarizes the scalability of the three runtimes under different concurrency, with QPS increased from 1 to 20. Across low-to-mid concurrency levels, \SystemName consistently remains a latency profile close to \textit{Reserved} while reducing end-to-end delay compared with \textit{On-demand}. At higher concurrency, the advantage over \textit{On-demand} remains substantial: at QPS=20, Laplace achieves 88.7s P99 E2E latency, a $2.9\times$ speedup over \textit{On-demand} (257.2s). This indicates that \SystemName can absorb increasing concurrency without inheriting the long-tail latency explosion of the \textit{On-demand} baseline.

Fig.~\ref{fig:e2e_scalability:cumulative} also reveals that the main scalability bottleneck of \textit{On-demand} lies in cumulative sandbox provisioning. In the low-QPS regime, the mean cumulative sandbox provisioning latency stays below $4$ seconds for all three modes. However, once the load reaches the higher-QPS regime ($\text{QPS}\geq 5$), \textit{On-demand} suffers growing degradation from network contention and resource constraints, and its cumulative sandbox provisioning latency rises steadily and eventually surpasses $50$ seconds. By contrast, \textit{Laplace} remains tightly bounded in the sub-$5$-second range and continues to slightly outperform \textit{Reserved} in provisioning latency, reflecting the benefit of \SystemName's intent-aware sandbox prewarming and stochastic sandbox prefetching under concurrency pressure.

\subsubsection{Resource Footprint and Efficiency}
\label{sec:exp:e2e_resource}


Fig.~\ref{fig:e2e_resource:cpu} shows that the CPU footprint of the three runtimes remains relatively compact under low-load conditions but diverges significantly as concurrency scales. While \textit{\SystemName} exhibits no obvious advantages under low QPS loads, its strengths gradually emerge in the high-QPS regime where \(\text{QPS}\geq 8\). It consistently restrains peak CPU utilization across all tested loads, with the peak resource consumption capped at around $12.2$ cores. This translates to a $22.8\%$--$23.3\%$ reduction in peak CPU pressure compared to both \textit{On-demand} and \textit{Reserved}, which demand roughly $15.8$--$15.9$ cores. Particularly in the high-QPS regime, \textit{\SystemName} stabilizes within a tight band of $\sim 11$--$12$ cores, whereas the two baselines fluctuate at a higher plateau of $14$--$16$ cores, demonstrating that \SystemName effectively mitigates CPU contention (saving up to $25\%$ CPU resource under peak load) without sacrificing sub-millisecond control-plane responsiveness.

The memory footprint (Fig.~\ref{fig:e2e_resource:memory}) exhibits a pronounced resource separation across the evaluation spectrum. As expected, \textit{Reserved} is the most memory-intensive variant due to its persistence strategy; its memory consumption surges from an initial $24.6$\,GiB at light load to a massive peak of $80.6$\,GiB, staying sustained above $60$\,GiB as concurrency intensifies. Conversely, \textit{On-demand} maintains a lean profile, bounding its peak usage between $24.3$\,GiB and $40.4$\,GiB. Notably, \textit{\SystemName} closely mirrors the trajectory of \textit{On-demand} across the majority of QPS configurations, topping out at $49.4$\,GiB. This represents a $45.9\%$ reduction in peak memory footprint compared to \textit{Reserved}, demonstrating that \SystemName successfully eliminates the prohibitive memory holding costs of long-run sandboxes, while maintaining a predictable, serverless-like resource elasticity under heavy concurrent workflows.

\subsection{Micro-Benchmarking}
\label{sec:exp:ablation}

This section provides mechanism-level attribution for the end-to-end improvements reported above. In particular, we isolate the effects of intent-aware sandbox prewarming, stochastic sandbox prefetching, semantic caching, and out-of-band data transmission to avoid conflating component contributions in the E2E section.

\subsubsection{Effectiveness of Intent-Aware Sandbox Prewarming}

We further decompose intent-aware sandbox prewarming into three orthogonal factors: the keyword trigger threshold, the semantic model family, and the hybrid fusion policy. All results are measured on the same sampled tasks and repeated 100 times.

\paragraph{Sensitivity of Keyword Router Threshold $\gamma$.}
To rigorously isolate the impact of the keyword matching threshold, we evaluate $\gamma\in\{1,2,3\}$ under identical workload profiles, executing $100$ independent 
trials for each configuration to ensure statistical convergence. As quantified in Table~\ref{tab:keyword_k_tradeoff}, $\gamma=2$ emerges as the optimal configuration for the Keyword Router. Specifically, $\gamma=2$ substantially compresses the average keyword-driven waiting latency to $323.0$\,ms (a $2.43\times$ 
reduction compared to $786.6$\,ms under $\gamma=1$) while maintaining a high predictive routing coverage of $95\%$. 

The severe latency degradation under $\gamma=1$ is primarily driven by its hyper-sensitivity, which triggers $20$ distinct tool mismatch instances across the $100$ runs; 
these false positives inadvertently amplify cold-start fallback penalties and introduce heavy tail latencies. Conversely, while increasing the threshold to $\gamma=3$ 
further sharpens routing precision (yielding a $97\%$ match rate with only $3$ mismatch instances), it severely erodes the prewarm lead time, causing the average 
waiting time to climb back to $621.8$\,ms. This empirical trade-off confirms that $\gamma=2$ effectively balances early predictive agility with resource stability.

\begin{table}[t]
\caption{Keyword threshold sensitivity ($\gamma=1,2,3$).}
\label{tab:keyword_k_tradeoff}
\centering
\small
\begin{tabular}{lccc}
\toprule
$\gamma$ & Avg Wait (ms) & Match Rate & Mismatch Runs \\
\midrule
1 & 786.619 & 80.0\% & 20 \\
2 & 323.021 & 95.0\% & 5 \\
3 & 621.812 & 97.0\% & 3 \\
\bottomrule
\end{tabular}
\vspace{-4mm}
\end{table}

\paragraph{Sensitivity of Semantic Router Models.}
Table~\ref{tab:semantic_model_compare} compares the performance and computational trade-offs of three semantic routing models under their optimal configurations:
\begin{itemize}[leftmargin=1.2em]
    \item \textbf{{Retrieval}:} This model is implemented as a non-neural, sparse token-level retriever. It leverages TF-IDF~\cite{sparck1972statistical} weighting combined with a $Top\text{-}N$ nearest neighbor aggregation network, constructing composite vector features across token unigrams, selective bigrams, and character $n$-grams (ranging from $3$ to $5$ characters).
    \item \textbf{{Encoder}:} This neural model is built upon a fine-tuned \textit{all-MiniLM-L6-v2}~\cite{encoder} transformer sequence representation network. It maps fluid multi-turn trajectories into dense vector spaces via contrastive pair-wise optimization under a \textit{CosineSimilarityLoss} constraint.
    \item \textbf{FastText:} This lightweight model~\cite{joulin2017bag} represents text as the average of word and subword embeddings followed by a linear classifier.
\end{itemize}

Empirically, as quantified in Table~\ref{tab:semantic_model_compare}, both \textit{Retrieval} and \textit{Encoder} models achieve a optimal prewarming Hit Rate, demonstrating that both sparse lexical tokens and dense semantic features can successfully capture all necessary tool invocation targets. However, \textit{Retrieval} achieves higher orchestration quality and routing precision, yielding a Micro-F1 of $0.970$ and a Precision of $0.942$, whereas \textit{Encoder} degrades to a Micro-F1 of $0.776$ and a Precision of $0.634$. 

Furthermore, from an runtime efficiency perspective, \textit{Retrieval} operates with a significantly lower computational footprint, executing with an average inference latency of only $2.116$\,ms--yielding a $3.70\times$ speedup compared to the $7.822$\,ms latency incurred by \textit{Encoder}. Conversely, while \textit{FastText} provides ultra-low execution latency ($0.171$\,ms), its shallow token representation space fails to deliver usable discrimination under multi-turn reasoning context drifts, resulting in a Micro-F1 and Hit Rate of merely $0.124$. Consequently, \textit{Retrieval} is selected as the production instance for Laplace's semantic branch, as it strikes the optimal Pareto-efficiency frontier 
between high-fidelity prediction accuracy and low-overhead control-plane latency.

\begin{table}[t]
\caption{Semantic router model comparison (best operating point per model).}
\label{tab:semantic_model_compare}
\centering
\small
\begin{tabular}{lcccc}
\toprule
Model & Micro-F1 & Hit Rate (Top-3) & Precision & Avg Latency (ms) \\
\midrule
Retrieval & 0.970 & 0.992 & 0.942 & 2.116 \\
Encoder & 0.776 & 0.968 & 0.634 & 7.822 \\
FastText & 0.124 & 0.124 & 0.124 & 0.171 \\
\bottomrule
\end{tabular}
\vspace{-4mm}
\end{table}

\paragraph{Sensitivity of Assembly Policies.}
To rigorously isolate the algorithmic impact of the intent combination layer, we evaluate three distinct fusion policy paradigms under identical multi-turn 
context distributions:
\begin{itemize}[leftmargin=1.2em]
    \item \textbf{Union ($\cup$):} The predictive prewarming primitive is non-blocking and asynchronously dispatched the exact microsecond either the keyword router or the semantic router hits their respective individual activation boundaries.
    \item \textbf{Intersection ($\cap$):} Predictive orchestration strictly enforces a dual-router consensus; a sandbox is prewarmed if and only if both the keyword router and semantic router concurrently validate the sandbox candidate's invocation intent.
    \item \textbf{Weighted:} This hybrid policy aggregates multi-modal intent metrics into a centralized candidate score: $s(c) = w_k \cdot s_k(c) + w_s \cdot s_s(c)$, 
    where $s_k(c)$ denotes the matched-keyword ratio (normalized against the static lexicon scale) and $s_s(c)$ represents the real-time semantic retrieval similarity vector. A predictive trigger is dispatched only when clearing a rigid threshold: $s(c) \geq \tau_f$. In our empirical runner, we implement a symmetric baseline configuration with $w_k = 0.5, w_s = 0.5$, and $\tau_f = 0.5$.
\end{itemize}

As quantified in Table~\ref{tab:hybrid_fusion_tradeoff}, there is a clear trade-off between the aggressive \textit{Union Assembly} and the conservative \textit{Intersection} policy. \textit{Union Assembly} prioritizes latency-masking agility, reducing the average waiting latency to just $124.45$\,ms at the cost of a slight $5.0\%$ cold-start ratio. In contrast, the conservative \textit{Intersection} policy eliminates cold starts ($0.0\%$) by enforcing strict dual-router consensus, but at the cost of blocking the critical path and driving the average latency up to $393.52$\,ms ($3.16\times$ higher than Union). 

The \textit{Weighted} policy performs the worst, inflating latency to $1308.38$\,ms due to a $45.0\%$ cold-start ratio. This failure stems from the structural misalignment between keyword tokens and high-dimensional semantics. Without dynamic re-normalization, shifting context distributions under multi-turn reasoning cause the combined scores to drift, frequently failing to clear the static $\tau_f = 0.5$ threshold. These results validate our choice of a \textit{Union-first} assembly design to maximize latency-masking performance while maintaining practical routing correctness.

\begin{table}[t]
\caption{Assembly policy trade-off.}
\label{tab:hybrid_fusion_tradeoff}
\centering
\small
\begin{tabular}{lccc}
\toprule
Assembly Mode & Avg Wait (ms) & Target Match Rate & Cold Start Ratio \\
\midrule
Union & 124.446 & 95.0\% & 5.0\% \\
Intersection & 393.523 & 100.0\% & 0.0\% \\
Weighted & 1308.378 & 55.0\% & 45.0\% \\
\bottomrule
\end{tabular}
\vspace{-4mm}
\end{table}

\subsubsection{Effectiveness of Stochastic Sandbox Prefetching}

\begin{figure}[t]
\centering
\includegraphics[width=0.75\linewidth]{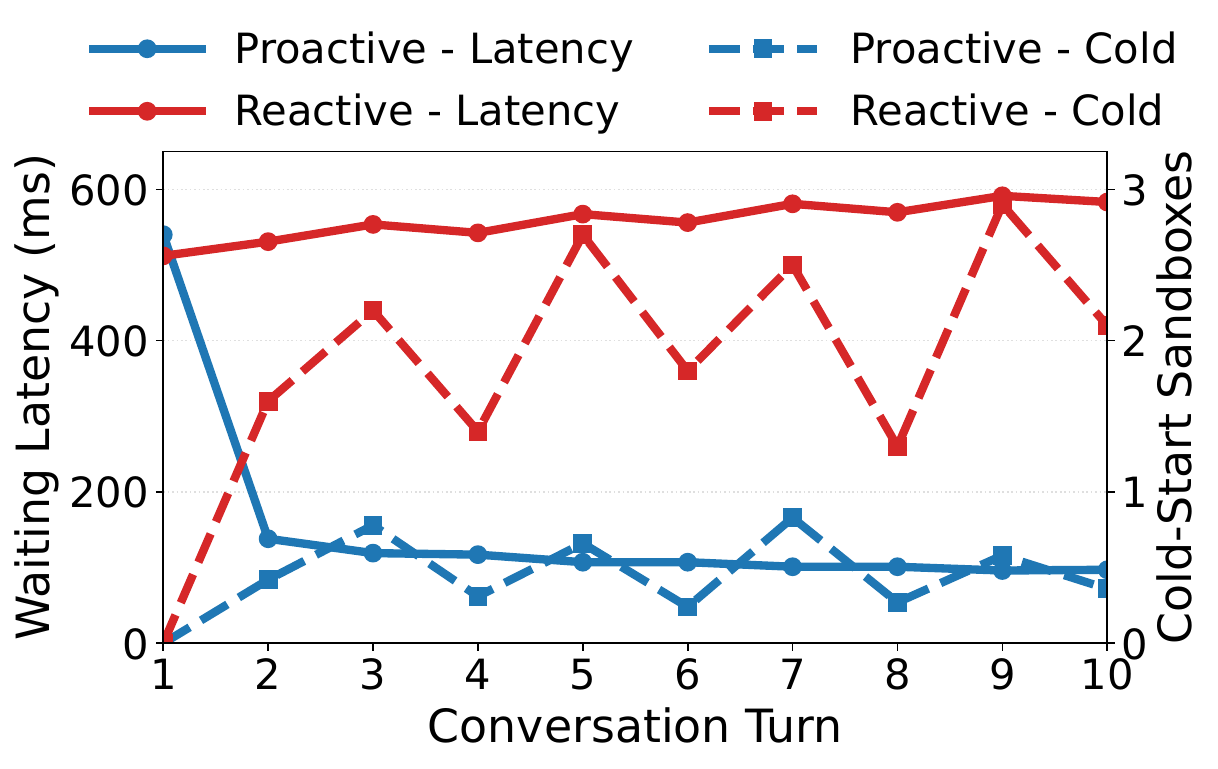}
\vspace{-0.3em}
\caption{Dynamic execution profiling across a 10-turn conversation horizon, benchmarking per-turn average waiting latency (left) against the corresponding average cold-start 
sandbox activation count (right).}
\label{fig:prefetch_turn_trend}
\vspace{-4mm}
\end{figure}

To isolate the contribution of stochastic prefetching under multi-turn agent execution, we evaluate Laplace with two deployment variants under an identical planning workload:
\begin{itemize}[leftmargin=1.2em]
    \item \textbf{\SystemName-Reactive:} A routing-only baseline that performs inline token-level intent detection but does not use cross-step transition prediction.
    \item \textbf{\SystemName-Proactive:} The full design that couples online hybrid routing with a stochastic Markovian prefetcher over the sandbox dependency graph (SDG), 
    enabling asynchronous sandbox preparation before the next step.
\end{itemize}

Fig.~\ref{fig:prefetch_turn_trend} reports per-turn average waiting latency and cold-start counts over a 10-turn horizon. During Turn 1, both variants show similar latency 
($512.06$\,ms vs. $540.06$\,ms), since no historical transition signal is available to initialize the Markov predictor. From Turn 2 onward, the two trajectories diverge sharply. 
\textit{\SystemName-Proactive} reduces average waiting latency from $540.06$\,ms to $138.14$\,ms at Turn 2 and further to $97.14$\,ms by Turn 10, while \textit{\SystemName-Reactive} 
increases to $583.26$\,ms at Turn 10. This corresponds to a $6.0\times$ end-horizon latency reduction. The mechanism is consistent with the cold-start telemetry. Under 
\textit{\SystemName-Reactive}, average cold-start count peaks at $2.90$ instances per turn (Turn 9), indicating repeated exposure 
to sandbox initialization cost on the critical path. In contrast, \textit{\SystemName-Proactive} keeps per-turn cold starts within $0.24$--$0.83$, effectively masking startup 
latency through early scheduling. These results show that Laplace improves multi-turn responsiveness through accurate, low-overhead temporal prefetching rather than 
infrastructure over-provisioning.

\subsubsection{Effectiveness of Semantic Cache}
\label{sec:exp:semantic_cache}

To quantify the impact of our semantic cache, we benchmark three variants under the same repeated-request workload over 100 runs:
\begin{itemize}[leftmargin=1.2em]
    \item \textbf{No Cache:} that executes every repeated request from scratch.
    \item \textbf{Exact-Match Cache:} A strict cache that matches on tool identity and normalized argument equality.
    \item \textbf{Semantic Cache:} reusing results when the tool identity matches and semantic similarity exceeds the threshold.
\end{itemize}

\begin{table}[t]
\caption{Semantic cache trade-off.}
\label{tab:semantic_cache}
\centering
\small
\begin{tabular}{lccc}
\toprule
Setting & Avg Wait (ms) & Hit Rate & Bypass Ratio \\
\midrule
No Cache & 412.78 & 0.0\% & 0.0\% \\
Exact-Match Cache & 233.64 & 33.6\% & 100.0\% \\
Semantic Cache ($\tau=0.6$) & 141.93 & 37.4\% & 84.8\% \\
\bottomrule
\end{tabular}
\vspace{-4mm}
\end{table}

As shown in Table~\ref{tab:semantic_cache}, while exact-match caching improves performance over \textit{No Cache}, semantic caching captures a wider envelope of near-duplicate requests by tolerating surface-form variations. Compared to \textit{No Cache}, semantic caching achieves a 2.91$\times$ speedup in average waiting latency (dropping from 412.78\,ms to 141.93\,ms) and increases the cache hit rate to 37.4\% . 

However, sematic caching is slightly lower than exact matching because a small fraction of semantic hits are not sufficiently reliable to skip sandbox setup and must fall back to normal execution after validation. This behavior is consistent with the intended design: semantic equivalence expands the reusable request space, but some loose matches still require conservative verification before execution can be bypassed.

\subsubsection{Effectiveness of Out-of-Band Data Transmission}
\label{sec:exp:oob}

\begin{figure}[t]
\centering
\includegraphics[width=0.75\linewidth]{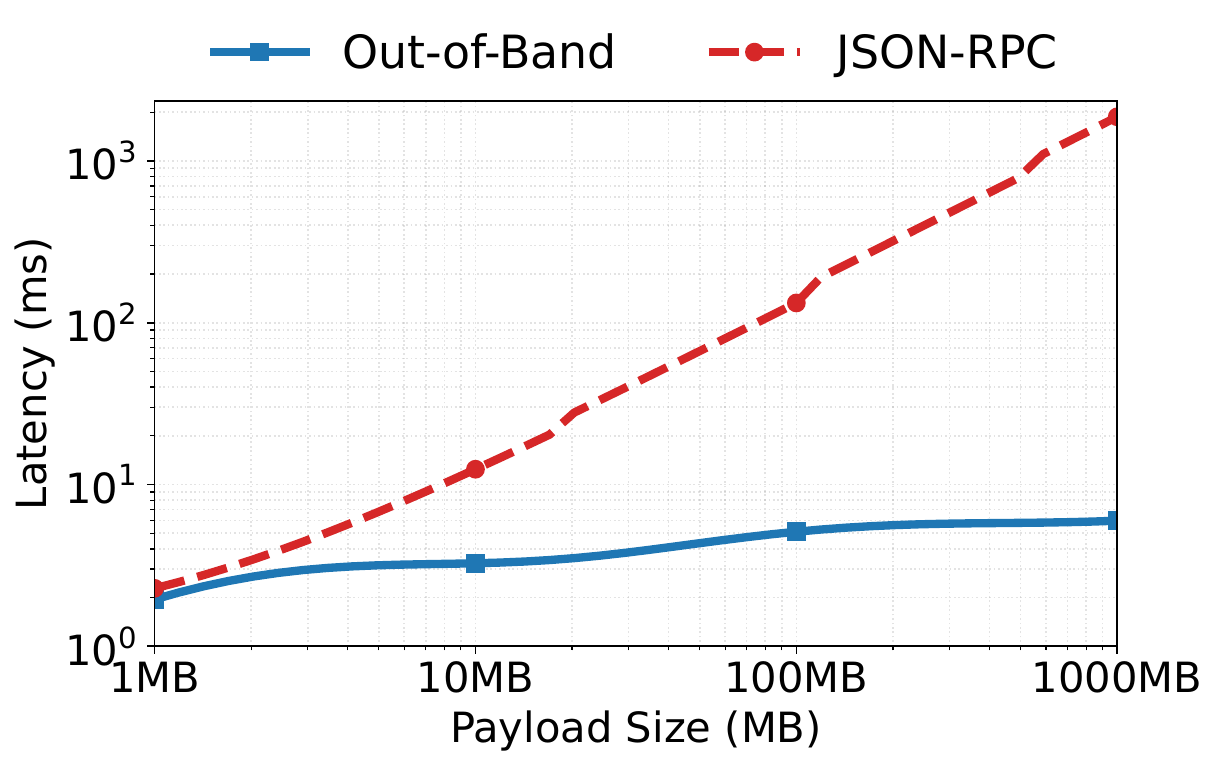}
\caption{Data transmission latency scaling profiles under varying payload sizes, comparing \SystemName's out-of-band data transmission against standard JSON-RPC serialization.}
\label{fig:oob_payload_latency}
\vspace{-4mm}
\end{figure}

To evaluate the architectural efficiency of our control-and-data plane separation, we isolate the data transmission overhead by benchmarking two distinct transport paradigms across an exponential data payload spectrum scaling from $1.00$\,MB to $1000.00$\,MB:
\begin{itemize}[leftmargin=1.2em]
    \item \textbf{Out-of-Band:} Our proposed out-of-band transport mechanism. It completely bypasses the control-plane RPC tunnel by writing bulky state payloads 
    directly to a dedicated local shared-memory substrate or zero-copy virtualized host-guest ring buffers, passing only lightweight, fixed-size references over the wire.
    \item \textbf{JSON-RPC:} The standard baseline paradigm utilized in conventional agent runtimes. It marshals multi-modal data payloads directly into the inline runtime execution stream, forcing the control-plane to serialize and transport raw data matrices synchronously via text-based JSON-RPC network primitives.
\end{itemize}

As shown in Fig.~\ref{fig:oob_payload_latency}, the empirical measurements reveal a clear scaling divergence between the two protocols across an exponential payload spectrum. At the baseline threshold ($1.00$\,MB), both configurations demonstrate sub-3 millisecond performance, with \textit{Out-of-Band} maintaining a slight edge ($1.95$\,ms vs. $2.28$\,ms). However, as the payload expands, the \textit{JSON-RPC} pipeline suffers a catastrophic linear performance degeneration; throttled by heavy serialization bottlenecks, its latency grows to $132.45$\,ms at $100.00$\,MB and reaches $1873.16$\,ms at the $1000.00$\,MB boundary. Conversely, \textit{Out-of-Band} demonstrates a near-constant $O(1)$ scaling profile, drifting to only $5.97$\,ms at the $1$\,GB boundary—a $313.55\times$ latency reduction.

This performance gap stems from eliminating critical-path serialization and memory copying. In multi-turn workflows, agents frequently exchange large multi-modal states like high-dimensional vectors or media files. Standard JSON-RPC stalls the control loop due to synchronous serialization, whereas \SystemName's out-of-band design decouples control signals from raw payload routing, reducing data transfer to a constant-time reference-passing operation. These results demonstrate that separating the control and data planes keeps \SystemName's orchestration overhead minimal and independent of payload size.
\section{Discussion}
\label{sec:discussion}

\cparagraph{Architectural Overhead} A critical concern in predictive serving runtimes is whether the system-level orchestration mechanisms introduce non-negligible processing penalties onto the critical path. In \SystemName, this control-plane overhead is thoroughly isolated from the GPU-bound inference loop through strict out-of-band execution and parallel routing mechanics. While the token-level scanning runs with deterministic $\mathcal{O}(1)$ complexity, the inherently heavier semantic retrieval engine is offloaded to dedicated background CPU worker threads. Because the intent router utilizes an \textit{Union Assembly} ($\cup$), the critical path never blocks for late-arriving semantic embeddings; any early fast-path trigger instantly dispatches the activation primitive within microseconds. Combined with the prefetching daemon--which evaluates low-dimension Markov matrix transitions bounded by the small cardinality of active sandboxes--\SystemName restricts its control-plane telemetry to tens of microseconds, securing a nearly zero-cost latency impact on token generation.

\cparagraph{Ecosystem Generalization} We position \SystemName as a runtime middleware layer operating between upstream reasoning agents and downstream execution environments. This design leverages two widely available capabilities in modern agent ecosystems~\cite{wu2023autogen, langgraph, gao2024agentscope}: streaming token-level outputs from upstream frameworks, and standardized tool interfaces provided by protocols such as MCP~\cite{mcp2024}. As a result, \SystemName requires no additional system-level constraints beyond what is already supported in existing agent and sandbox orchestration stacks.

Importantly, this execution abstraction extends beyond inference-time serving to agentic reinforcement learning (Agentic RL) frameworks, such as VERL~\cite{sheng2024hybridflow} and Slime~\cite{slime_github}, where rollout generation and environment interaction follow a structurally similar loop. In these Agentic RL settings, \SystemName’s runtime optimizations—namely \textit{intent-aware sandbox prewarming} and \textit{stochastic sandbox prefetching}—can be seamlessly adapted to mask environment initialization and interaction overheads during training rollouts with modest integration effort.

\cparagraph{Limitations and Future Work} Despite its performance gains, \SystemName exhibits certain architectural boundaries. On the control plane, our prefetcher assumes first-order history dependence via a Markov Chain model, which may suffer from diminished prediction accuracy during open-ended, long-horizon agent workflows. Crucially, \SystemName inherently mitigates this via its hierarchical, two-tiered preparation: any inter-step prefetch miss gracefully falls back to the intra-step intent-aware prewarming during real-time token streaming. This design guarantees that worst-case environment latency remains tightly bounded within the LLM's decoding phase. Future work will investigate online Graph Neural Networks (GNNs) to adaptively capture non-linear transition patterns.

On the data plane, our \texttt{mmap}-based transmission requires co-locating the Agent Engine and sandboxes within the same host boundary. While this design seamlessly aligns with mainstream multi-tenant serving topologies (e.g., Kubernetes Pod IPC sharing or Sidecar patterns) and leverages robust container-level namespace and cgroup security isolation, it restricts single-session cross-node scaling. To support large-scale distributed clusters, we plan to integrate RDMA-assisted zero-copy transmission, extending our out-of-band data plane into disaggregated cloud infrastructures.

\vspace{-0.2em}
\section{Related Work}

\cparagraph{Lightweight Sandbox Runtimes}
The systems community has actively explored lightweight isolation mechanisms to mitigate the physical instantiation cost of serverless execution environments. Notable advancements include micro-virtual machines (MicroVMs) like Firecracker~\cite{agache2020firecracker} and RunD~\cite{li2022rund}, WebAssembly (Wasm)~\cite{wasm2026} runtimes, and process-level snapshotting/forking frameworks such as FaaSnap~\cite{ao2022faasnap} and TrEnv-X~\cite{10.1145/3805475}. 
These infrastructure-level systems focus on minimizing localized host setup or exploiting hardware-assisted remote memory pools (e.g., CXL/RDMA) to share and reuse physical sandboxes across tenants. 

Crucially, these low-layer container optimizations are entirely orthogonal to \SystemName. While microVM snapshots or repurposable sandboxes successfully compress the physical infrastructure boot time to milliseconds, application-layer handshakes (such as MCP discovery hooks) and environment attachment bottlenecks still natively linger on the critical runtime path. \SystemName operates at a higher, application-perceptive orchestration layer; it complements these physical speedups by exploiting a distinct temporal dimension---overlapping control-plane scheduling with streaming token generation---to fully mask, rather than compress, the intrinsic readiness latency.

\cparagraph{Serverless Predictive Prewarming}
Predictive prewarming is a widely embraced technique in standard serverless frameworks to eradicate the notorious tail-latency penalties of tenant cold starts. State-of-the-art prefetching daemons (e.g., Mitosis~\cite{wei2023no} and IceBreaker~\cite{roy2022icebreaker}) primarily depend on historical time-series analytics, statistical invocation frequency histograms, or temporal correlation clustering to predictively prepare idle container runtimes. 

However, these classical prewarming paradigms inherently assume that incoming requests follow independent and identically distributed arrival models or deterministic time-triggered patterns. This core assumption breaks down completely under emerging LLM Agent workloads. In multi-turn autonomous agent reasoning loops, tool invocations are dynamically, autonomously, and non-linearly determined by the LLM’s fluid context trajectory, presenting complex execution dependency horizons. \SystemName bridges this gap by co-designing the serving infrastructure with explicit agent behavioral characteristics, introducing a stochastic Markovian predictive framework built over a sandbox dependency graph (SDG) to model real-time autonomous state transitions.

\cparagraph{LLM Serving and Agent Orchestration}
Accelerating the end-to-end execution of Large Language Models has driven extensive research across the AI systems spectrum. On the model-serving boundary, mainstream runtimes such as vLLM~\cite{kwon2023efficient} and SGLang~\cite{zheng2024sglang} optimize GPU-internal kernel execution, memory caching via PagedAttention, and automated speculative decoding. Parallel to this, application-level agent orchestration platforms like AgentScope~\cite{gao2024agentscope}, AutoGen~\cite{wu2023autogen} and LangGraph~\cite{langgraph} provide modular abstractions for programming multi-turn autonomous multi-agent teams. 

Nevertheless, existing LLM serving engines fundamentally treat model execution as a self-contained GPU computing unit, entirely oblivious to the physical host-plane environment friction when interacting with external tools. Conversely, high-level agent frameworks lack system-level visibility into cloud-native host topologies, resulting in uncoordinated data transfer and execution stalls. \SystemName operates as a runtime middleware layer that bridges this gap, enabling co-optimized out-of-band control-plane predictive routing and shared-memory data transmission across heterogeneous multi-tenant host execution boundaries.

\section{Conclusions}

We present \SystemName, a predictive execution runtime for LLM-based agent systems that reduces tail latency and resource inefficiency in multi-tenant environments. By analyzing the end-to-end agent execution step, we find that inefficiencies primarily arise from rigid dependencies among reasoning, environment initialization, and sandbox execution. \SystemName addresses this through a unified design that enables temporal overlap between LLM execution and sandbox setup, anticipates future sandbox needs across steps to reduce cross-step cold-start overheads, and eliminates redundant computation and communication through reuse-aware execution and out-of-band data transport. \SystemName is implemented on top of AgentScope~\cite{gao2024agentscope} and evaluated it under highly concurrent multi-turn workloads. Results show up to a 2.9$\times$ reduction in P99 latency and 45.9\% lower peak memory usage, with a 97.9\% prewarming hit rate. These results demonstrate that exploiting temporal overlap across the agent execution loop is an effective and general mechanism for improving efficiency in large-scale agent deployments.




\bibliographystyle{ACM-Reference-Format}
\bibliography{references}



\end{document}